\documentstyle[preprint,aps,epsf]{revtex}

\begin{document}
\draft
\title{Dual description of the superconducting phase transition}
\author{Michael Kiometzis, Hagen Kleinert, and Adriaan M.J. Schakel}
\address{Institut f\"ur Theoretische Physik \\
Freie Universit\"at Berlin \\
Arnimallee 14 \\
14195 Berlin}
\maketitle
\begin{abstract}
The dual approach to the Ginzburg-Landau theory of a Bardeen-Cooper-Schrieffer
superconductor is reviewed.  The dual theory describes a grand canonical
ensemble of fluctuating closed magnetic vortices, of arbitrary length and
shape, which interact with a massive vector field representing the local
magnetic induction.  When the critical temperature is approached from below,
the magnetic vortices proliferate.  This is signaled by the disorder field,
which describes the loop gas, developing a non-zero expectation value in the
normal conducting phase.  It thereby breaks a {\it global} U(1) symmetry.  The
ensuing Goldstone field is the magnetic scalar potential.  The
superconducting-to-normal phase transition is studied by applying
renormalization group theory to the dual formulation.  In the regime of a
second-order transition, the critical exponents are given by those of a
superfluid with a reversed temperature axis.
\end{abstract}
\section{Introduction}
\label{sec:intro}
In a recent letter \cite{KKS} critical indices of the superconducting-to-normal
phase transition were derived starting from a dual formulation of the
Ginzburg-Landau theory.  The purpose of the present article is to review this
approach.  The article, which is pedagogical in tone and attempts to be
self-contained, highlights also some recent developments related to the
symmetry aspects of the dual formulation which are important for understanding
the phase transition.

The basic idea of the dual approach originates from three-dimensional ($3D$)
lattice studies carried out more than 15 years ago \cite{BMK,Peskin,TS,Savit}.
These studies were instigated by the success of the
Berezinskii-Kosterlitz-Thouless (BKT) theory of the $2D$ $xy$ model developed
in the early seventies to describe the phase transition in a superfluid film
\cite{Be,KT73}.  It was argued that the phase transition has its origin in the
behavior of point-like vortices.  In the ordered low-temperature phase, pairs
of vortices with opposite topological charge are tightly bound.  Above the
critical temperature, these pairs unbind and the liberated vortices disorder
the system, turning the superfluid into a normal fluid.

The lattice studies of the pure $3D$ $xy$ model \cite{BMK} and its extension
to a lattice superconductor \cite{Peskin,TS} were aimed at obtaining an
analogous description of the phase transitions in these models in terms of
vortex {\it loops}.  A detailed account of these matters as well as an
extensive list of references to the literature can be found in the textbook by
Kleinert \cite{GFCM}.

Another development illuminating the dual approach to the Ginzburg-Landau
model was initiated in Ref.\ \cite{KRE}.  The basic observation was that since a
local gauge symmetry can never be broken \cite{Elitzur}, a
local gauge description of a phase transition is not feasible.  It was argued
that the $3D$ Ginzburg-Landau theory contains, in addition to the local gauge
symmetry, another {\it global} U(1) symmetry.  When considered in $2+1$
space-time dimensions with a Minkowski metric, this symmetry is generated by
the magnetic flux operator.  It was demonstrated that this
symmetry is broken in the normal phase, while it is unbroken in the
superconducting phase.  An order parameter was given, and it was shown that
the massless photon of the normal phase is the Goldstone particle associated
with the broken flux symmetry.

To clarify the concept of duality it is expedient to remain for the moment in
$2+1$ space-time dimensions.  There the theory is a quantum field theory
possessing particle states which are created by field operators.  This is to
be contrasted with the space-time approach of Feynman \cite{Feynman} where
worldlines of particles are the fundamental objects.  The particle's mass is
given by the tension of the worldline.  Now, suppose that a quantum field
theory contains point-like solitons, i.e., topologically stable,
time-independent solutions of the field equations.  When set into motion such
a soliton traces out a non-trivial worldline in space-time.  An operator
description of these excitations in terms of the original fields is non-local
and quite difficult.  The aim of the dual approach of the type we are
considering is to describe them by a local field, called a disorder field
\cite{GFCM}.  The space-time description of the particle is thus converted
into an equivalent field-theoretic one.

In the context of the (three dimensional Euclidean) Ginzburg-Landau theory the
notion of a particle state is of course missing.  The above picture has,
however, a direct counterpart, viz., worldlines of the space-time approach
correspond to vortices in the Euclidean theory.  More precisely, in the
setting of statistical (as opposed to quantum) field theory, a dual theory
describes vortices by a disorder field.  (In this discussion we restricted the
notion of duality to line defects, thereby omitting dual theories of other
topological defects such as domain walls and---in the context of quantum field
theory---instantons.)

The advantage of the dual theory is that it allows for a simple description of
the superconducting phase transition in the sense of Landau which is
impossible in the original formulation, as we shall discuss.  The order
parameter of the transition is the field representing the magnetic vortices.
This Landau theory, involving only a global symmetry and a genuine order
parameter, can---as usual---be taken as a convenient starting point of a
renormalization group analysis of the superconducting phase transition
\cite{KKS}.

The layout of the paper is as follows.  In the next section we illustrate the
dual approach by considering the Ginzburg-Landau model in two space
dimensions.  This example gives us important clues towards the form of the
dual formulation in $3D$.  In Sec.\ \ref{sec:twofold} we review the fact that
a loop gas can be succinctly described near a critical point by a $|\psi|^4$
field theory.  This equivalence plays a central role in the subsequent part of
the paper.  In Sec.\ \ref{sec:3DL} we discuss the dual formulation of a $3D$
lattice superconductor, and in Sec.\ \ref{sec:dual} we review the formulation
in the continuum.  In Sec.\ \ref{sec:mixed} we argue that the dual theory
yields the standard description of the mixed state of a type-II
superconductor.  As a consistency check we consider in Sec.\ \ref{sec:normal}
the dual formulation of the dual theory and demonstrate that the resulting
theory is the original Ginzburg-Landau model.  In Sec.\ \ref{sec:sop} we
discuss the gauge-invariant superconducting order parameter.  Finally, in
Sec.\ \ref{sec:Landau} we apply renormalization group theory to the dual
theory to show that the superconducting phase transition is of second order.
The critical exponents are shown to be in the universality class of the $xy$
model with the temperature axis reversed.
\section{Dual map of a 2$D$ superconductor}
\label{sec:2D}
Before discussing the theory of a superconductor in three space dimensions, it
is instructive to first study the two-dimensional case.  This exercise will
exhibit the relevant variables in a dual description and show what in
principle can be expected from such a description.

The Ginzburg-Landau model resulting from the Bardeen-Cooper-Schrieffer (BCS)
theory of superconductivity is specified by the Hamiltonian
\begin{equation}
\label{GL}
H = \int d^2 x \left[ {1 \over 2} (\nabla \times {\bf A})^2 + |(\nabla -2 i e
{\bf A})\phi|^2 + m_\phi^2 |\phi|^2 + \lambda  |\phi|^4 \right],
\end{equation}
with ${\bf A}$ the electromagnetic gauge potential.  We adopt the standard
notation and denote by $\nabla \times {\bf A} = \epsilon_{i j} \partial_i A_j$
the local induction.  The symbol $B$ is reserved for the spatial average
of the local field, and will be referred to as the magnetic induction, or flux
density.  These quantities have only one component in two space dimensions,
pointing in the missing third direction.  The coupling $2e$ accounts for the
double charge of Cooper pairs.  The complex scalar field $\phi$ is the
superconducting order field with a mass parameter $m_\phi$ and a self-coupling
$\lambda$.  In the superconducting phase of the model, $m_\phi^2 < 0$.

A useful limit of the theory is the so-called London limit where $m_\phi^2
\rightarrow -\infty$ and $\phi$ can be written as $\phi (x) = \exp[i\theta(x)]
w/\sqrt{2}$, with a constant $w$.  This limit properly represents many
relevant properties of the superconducting state.  The functional integral
over the size fluctuations of the scalar field can in this limit be
approximated by the saddle point, and $H$ becomes, omitting irrelevant terms
\begin{equation}
\label{hydroenergy}
H = \int d^2 x \left[ \frac{w^2}{2} {\bf p}_{\rm s}^2 +
\frac{1}{2} (\nabla \times {\bf A})^2 \right],
\end{equation}
with ${\bf p}_{\rm s}$ the superfluid momentum
\begin{equation}
\label{velo}
{\bf p}_{\rm s} = \nabla \theta - 2e {\bf A} +
\bbox{\theta}^{\rm P}.
\end{equation}
A so-called plastic field $\bbox{\theta}^{\rm P}$ must be introduced for
the following reason.  The variable $\theta$ is a phase variable and
is thus cyclic with periodicity $2\pi$.  A jump in $\theta$ by $2\pi$
along a cut in the spatial plane indicates the presence of a point
vortex.  When this is the case the Hamiltonian requires a field
$\bbox{\theta}^{\rm P}$ to compensate for these jumps so that $H$
remains smooth.  Such a construction with a compensating field is well
known in the Villain formulation of lattice models, and it can also be
given in the continuum \cite{GFCM}.  The curl of $\bbox{\theta}^{\rm
P}$ gives the density of vortices consisting of $\delta$ functions at
each vortex position
\begin{equation}
\label{vdensity}
\nabla \times \bbox{\theta}^{\rm P} = 2\pi
\sum_{\alpha} n_{\alpha} \delta({\bf x} - {\bf x}_{\alpha}),
\end{equation}
where $n_{\alpha}$ is the winding number of a vortex, and ${\bf
x}_{\alpha}$ its location.  We will restrict ourselves to vortices
with unit winding number, $n_{\alpha} = \pm 1$ for a vortex and
antivortex, respectively.  Since the theory is charged, vortices carry
also a magnetic flux quantum $n_{\alpha} \, \pi/e$, and are more
properly called {\it magnetic vortices}.

The solution of (\ref{vdensity}) is given by \cite{GFCM}
\begin{equation}
\label{line}
\theta_i^{\rm P} = - 2 \pi \sum_{\alpha} n_{\alpha} \epsilon _{ij} \,
\delta_j({\bf x},L_{\alpha}).
\end{equation}
where $\delta_i({\bf x},L_{\alpha})$ is the $\delta$ function on the
line $L_{\alpha}$ which starts at the position of the $\alpha$th
vortex ${\bf x}_{\alpha}$ and runs to spatial infinity along
an arbitrary path:
\begin{equation}
\delta_i({\bf x},L_{\alpha}) = \int_{{\bf x}_{\alpha}}^{\infty} d y_i
\, \delta({\bf x} - {\bf y}).
\end{equation}
With the help of Stokes' theorem in two dimensions which can be
expressed in differential form as $\nabla \cdot \bbox{\delta}({\bf
x},L_{\alpha}) = \delta({\bf x} - {\bf x}_{\alpha})$ it is easy to
check that the explicit expression (\ref{line}) for the field
$\theta_i^{\rm P}$ indeed solves Eq.\ (\ref{vdensity}).  The shape of
the line $L_{\alpha}$ is physically irrelevant, only the starting
point which marks the position of the vortex matters.  Leaving this
point fixed the shape of $L_{\alpha}$ may be changed at will by
performing the transformation
\begin{equation}
\label{vgt}
\theta_i^{\rm P}({\bf x}) \rightarrow \theta_i^{\rm P} ({\bf x}) +
\partial _i \delta ({\bf x},S),
\end{equation}
where $\delta ({\bf x},S)$ is the $\delta$ function on the surface $S$
swept out by $L_{\alpha}$:
\begin{equation}
\delta ({\bf x},S) := \int_S d\sigma d\tau \, \epsilon_{i j} \,
\partial_{\sigma} y_i \partial_{\tau} y_j \, \delta[{\bf
x} - {\bf y} (\sigma ,\tau )],
\end{equation}
with ${\bf y} (\sigma, \tau )$ being a parameterization of $S$.  The
superfluid momentum and thus the Hamiltonian is invariant under these
so-called vortex gauge transformations \cite{GFCM}.

For the time being vortices are ignored by setting the plastic field
$\bbox{\theta}^{\rm P}$ to zero.  The partition function of the system
is then given by
\begin{equation}
\label{znovor}
Z = \int {\cal D}\theta \int {\cal D}{\bf A} \, \Xi ({\bf A})
\, {\rm e}^{-H},
\end{equation}
with $\Xi({\bf A})$ a gauge-fixing factor for the gauge field ${\bf A}$, and
$H$ the Hamiltonian (\ref{hydroenergy}).  Here, fields and coupling constants
are rescaled so that no explicit temperature dependence appears in the
Boltzmann factor.  In (\ref{znovor}) it is easy to integrate out the $\theta$
fluctuations.  This yields for the partition function
\begin{equation}
\label{schwlike}
Z = \int {\cal D}{\bf A} \, \Xi({\bf A}) \, {\rm
exp}\left\{-\frac{1}{2} \int d^2 x \left[(\nabla \times {\bf A})^2 +
m_A^2 A_i \left( \delta_{i j} - \frac{\partial _i
\partial_j}{\nabla^2} \right) A_j \right] \right\},
\end{equation}
where the last term with $m_A = 2 e w$ is a gauge-invariant, albeit non-local
mass term for the gauge field generated by the Higgs mechanism.  A massless
gauge field in $2D$ represents no physical degrees of freedom.  In Minkowski
space-time, this is easily understood by recognizing that in $1+1$ dimensions
there is no transverse direction.  The system contains therefore only a single
physical degree of freedom before the Higgs mechanism took place, namely
$\theta$.  This equals the number afterwards since a massive vector field
represents only one independent degree of freedom in $2D$.  Note that the
absence of genuine long-range order in $2D$ is no obstacle to the Higgs
mechanism.  (The massless Schwinger model \cite{Schwingermodel} provides the
simplest soluble example for this.)

We next introduce an auxiliary field ${\sf h}$ to linearize the first term in
(\ref{schwlike}),
\begin{equation}
\label{efield}
\exp \left[-\frac{1}{2} \int d^2 x  (\nabla \times {\bf A})^2
\right] = \int {\cal D}{\sf h} \, {\rm exp}\left[-\frac{1}{2} \int d^2
x \, {\sf h}^2 + i \int d^2 x \, {\sf h} \cdot (\nabla
\times {\bf A}) \right],
\end{equation}
and integrate out the gauge-field fluctuations [after adding a
gauge-fixing term $(1/2\zeta)(\nabla \cdot {\bf A})^2$].  The result is a
manifestly gauge-invariant expression for the partition function in
terms of a massive scalar field ${\sf h}$, which represents the single
degree of freedom contained in the theory:
\begin{equation}
\label{massivescalar}
Z = \int {\cal D} {\sf h} \, {\rm exp}\left\{-\frac{1}{2} \int d^2 x
\left[ \frac{1}{m_A^2} (\nabla {\sf h})^2 + {\sf h}^2 \right] \right\}.
\end{equation}
To understand the physical significance of the ${\sf h}$ field
appearing in this functional integral, we note that it follows from
(\ref{efield}) that the auxiliary field ${\sf h}$ satisfies the
equation
\begin{equation}
\label{id}
{\sf h} = i \nabla \times {\bf A}.
\end{equation}
That is, the fluctuating field ${\sf h}$ coincides up to a factor $i$ with the
local induction.  Equation (\ref{massivescalar}) shows that the magnetic
field has a finite penetration depth $\lambda = 1/m_A$.  In contrast to the
Ginzburg-Landau description where the functional integral runs over the gauge
potential, the integration variable in (\ref{massivescalar}) is the physical
field.

We now include vortex contributions.  The mass $m_A$ provides the system with
an infrared cutoff so that a single vortex in the charged theory has a finite
energy, implying that there will always exist thermally activated vortices.
This is different from the neutral model, describing a $2D$ superfluid, where
the absence of an infrared cutoff permits only tightly bound vortex-antivortex
pairs to exist.  We expect, accordingly, the superconducting phase to describe
a plasma of vortices, each carrying one magnetic flux quantum $\pm \pi/e$.  The
partition function now reads
\begin{equation}
\label{vincluded}
Z = \sum_{n_{+},n_{-}=0}^{\infty} \frac{z^{n_{+}+n_{-}}}{n_{+}!\,
n_{-}!} \int \prod_{\alpha} d^2 x_{\alpha} \, \int {\cal D}\theta
\int {\cal D}{\bf A} \, {\rm exp}\left\{-\frac{1}{2} \int d^2 x \,
[w^2 {\bf p}_{\rm s}^2 + (\nabla \times {\bf A})^2] \right\},
\end{equation}
where the superfluid momentum ${\bf p}_{\rm s}$ contains the vortex gauge
field $\bbox{\theta}^{\rm P}$, as in (\ref{velo}).  In (\ref{vincluded}),
$\int \prod_{\alpha} d^2 x_{\alpha}$ denotes the integration over the
positions of the vortices and the factor $1/n_{+}! \; (1/n_{-}!)$ accounts for
the fact that the (anti-)vortices are indistinguishable. The fugacity $z$ is
the Boltzmann factor for an isolated vortex $z = \exp(-\epsilon_{\rm c})$,
with $\epsilon_{\rm c}$ being the vortex core energy.

The field $\bbox{\theta}^{\rm P}$ can be shifted from the first to the second
term in (\ref{vincluded}) by applying the transformation ${\bf A} \rightarrow
{\bf A} - \bbox{\theta}^{\rm P}/(2e)$.  This results in the shift
\begin{equation}
\nabla \times {\bf A} \rightarrow \nabla \times {\bf A} - B^{\rm P},
\end{equation}
with the plastic field
\begin{equation}
\label{BP}
B^{\rm P} = -\frac{\pi}{e} \sum_{\alpha}
n_{\alpha} \, \delta({\bf x} - {\bf x}_{\alpha })
\end{equation}
representing the magnetic flux density.  Repeating the steps of the
previous paragraph we now obtain instead of (\ref{massivescalar})
\begin{equation}
\label{vortexsum}
Z = \sum_{n_{+},n_{-}=0}^{\infty} \frac{z^{n_{+}+n_{-}}}{n_{+}!\,
n_{-}!}  \int \prod_{\alpha} d {\bf x}_{\alpha} \, \int {\cal D} {\sf
h} \, {\rm exp}\left\{-\frac{1}{2} \int d^2 x \left[ \frac{1}{m_A^2}
(\nabla {\sf h})^2 + {\sf h}^2 \right] + i \int d^2 x B^{\rm P}
{\sf h}\right\}.
\end{equation}
where ${\sf h}$ represents the physical local field
\begin{equation}
{\sf h} = i (\nabla \times {\bf A} - B^{\rm P}),
\end{equation}
as will be clarified in a later section (Sec.\ \ref{sec:dual}).  The last term
in (\ref{vortexsum}) shows that the charge $g$ with which a magnetic vortex
couples to the fluctuating ${\sf h}$ field is the product of an elementary
flux quantum $\pi/e$ (which is contained in the definition of $B^{\rm
P}$) and the inverse penetration depth $m_A$, i.e.,
\begin{equation}
\label{g}
g = \frac{\pi}{e} m_A.
\end{equation}
For small fugacities the summation indices $n_{+}$ and $n_{-}$ can be
restricted to the values $0,1$ and we arrive at the partition function of the
massive sine-Gordon model \cite{Schaposnik}
\begin{equation}
\label{sineGordon}
Z = \int {\cal D} {\sf h} \, {\rm exp} \left( - \int d^2 x
\left\{\frac{1}{2} \left[\frac{1}{m_A^2} (\nabla {\sf h})^2 + {\sf
h}^2\right]- 2N \cos \left( \frac{\pi}{e} {\sf h} \right) \right\}
\right).
\end{equation}
This is the dual formulation of a $2D$ superconductor.  The vortices of unit
winding number $n_\alpha = \pm 1$ turned the otherwise free theory
(\ref{massivescalar}) into an interacting one.

The final form (\ref{sineGordon}) demonstrates the basic concepts of a dual
formulation: it is a formulation directly in terms of a (gauge-invariant)
field representing the physical local induction.  This is different from the
Ginzburg-Landau description of a two-dimensional superconductor where the
magnetic field is the curl of an unphysical gauge potential ${\bf A}$.
The dual formulation also accounts for the topological excitations, viz., the
magnetic vortices which in $2D$ are point defects.  They are coupled with a
charge $g = (\pi/e) m_A$ to the magnetic field.

In three space dimensions these two basic ingredients will surface again.
Since in $3D$ the magnetic field has three components of which two are
independent, the dual formulation will involve a massive {\it vector} field,
rather than a simple massive scalar field ${\sf h}$ as in $2D$.  But what is
more important, the point vortices of the two dimensional case become line
defects in $3D$.  A grand canonical ensemble of closed loops requires, as will
be demonstrated in the next section, a fluctuating field of its own to specify
it.  That is, accounting for the vortices does not simply add an interaction
term to the theory without vortices, but adds a whole new theory.  The
coupling of this new theory---which turns out to be a complex $|\psi|^4$
theory---to the field representing the local field is again given by $(\pi/e)
m_A$ as in two dimensions.
\section{Functional-integral description of a loop gas}
\label{sec:twofold}
The subject of this section is the well-known fact that a loop gas can be
described with the help of a functional integral involving a complex
$|\psi|^4$ theory \cite{GFCM,Parisi}.  This will be used in a later section
when formulating the dual theory of the Ginzburg-Landau theory in $3D$.  This
theory features fluctuating closed magnetic vortices described by a disorder
field theory.  We will review here the derivation in the continuum, the
discussion of the lattice derivation is relegated to the Appendix
\cite{GFCM,AID}.

Our point of departure is the correlation function $G({\bf x})$ of a
free complex field theory defined by the Hamiltonian
\begin{equation}
\label{Hamilton}
H = \int d^3x \left(|\nabla \psi|^2 + m^2 |\psi|^2 \right).
\end{equation}
Explicitly,
\begin{equation}
G({\bf x}) = \int \frac{d^3 k}{(2 \pi)^3} \frac{{\rm e}^{i {\bf k}
\cdot {\bf x}}}{{\bf k}^2 + m^2}.
\end{equation}
This is written in the Schwinger proper-time representation as an integral
over the proper time $s$ \cite{proptime}
\begin{eqnarray}
\label{green}
G({\bf x}) &=& \int_0^{\infty} ds \, {\rm e}^{-s m^2} \int \frac{d^3
k}{(2 \pi)^3} {\rm e}^{i {\bf k} \cdot {\bf x} } {\rm
e}^{-s{\bf k}^2} \nonumber \\ &=& \int_0^{\infty} ds \, {\rm e}^{-s m^2}
\left( \frac{1}{4 \pi s} \right)^{3/2} {\rm
e}^{-\case{1}{4} {\bf x}^2/s},
\end{eqnarray}
where we used the identity
\begin{equation}
\label{gamma}
\frac{1}{a} = \int_0^\infty d s \, {\rm e}^{- s a}.
\end{equation}
According to Feynman's formulation of quantum mechanics \cite{Feynman} the
right-hand side of (\ref{green}) can be represented as a sum over all
real-space paths of a particle with mass $\case{1}{2}$ running from $0$ at
imaginary time 0 to ${\bf x}$ at time $s$:
\begin{equation}
\label{feynrep}
\left( \frac{1}{4 \pi s} \right)^{3/2} {\rm e}^{-\case{1}{4}{\bf x}^2/s}
= \int_{{\bf x}(0)=0}^{{\bf
x}(s)={\bf x}} {\cal D} {\bf x}(s') \exp \left[ - \case{1}{4}
\int_0^s ds' \, \dot{\bf x}^2 (s') \right],
\end{equation}
involving the free ``Lagrangian'' $L = \dot{\bf x}^2 /4$.  The extra
Boltzmann factor $\exp (-s m^2)$ in (\ref{green}) suppresses large
proper-time values exponentially, and the integral $\int_0^\infty d s$
shows that all positive values of $s$ are allowed and accounted for.

From integrating the identity (\ref{gamma}) with respect to $a$ it
follows that to within an additive constant
\begin{equation}
\ln (a) = - \int_0^\infty \frac{d s}{s} {\rm e}^{-s a}.
\end{equation}
Employing this we can use the previous result to write
\begin{eqnarray}
\label{oint}
{\rm Tr} \ln (- \nabla^2 + m^2) &=& - \int_0^{\infty}
\frac{ds}{s} {\rm e}^{-s m^2} \int \frac{d^3 k}{(2 \pi)^3} {\rm
e}^{-s {\bf k}^2} \nonumber \\ &=& - \int_0^\infty \frac{d s}{s} {\rm e}^{-s
m^2} \oint {\cal D} {\bf x} (s') \exp \left[ - \case{1}{4} \int_0^s ds' \,
\dot{\bf x}^2 (s') \right],
\end{eqnarray}
where Tr denotes the integral over the momentum variables
\begin{equation}
{\rm Tr} \ln (- \nabla^2 + m^2) = \int \frac{d^3 k}{(2 \pi)^3} \ln ({\bf k}^2
+ m^2).
\end{equation}
The path integral $\oint {\cal D} {\bf x} (s')$ in (\ref{oint}) runs over
closed loops starting and ending in $0$.  Using the identity
\begin{equation}
{\rm Tr} \ln (- \nabla^2 + m^2) = \ln {\rm Det} (- \nabla^2 + m^2),
\end{equation}
we can write the inverse determinant as
\begin{equation}
\label{det}
{\rm Det}^{-1}(-\nabla^2 + m^2) = \exp (-W_0),
\end{equation}
with $W_0$ denoting the right-hand side of (\ref{oint}).  The exponential
$\exp(-W_0)$ has the expansion
\begin{equation}
\label{explicit}
{\rm e}^{-W_0} = \sum_{N=0}^{\infty} \frac{1}{N!} \prod_{l=1}^N \left[
\int_0^\infty \frac{d s_l}{s_l} {\rm e}^{-m^2 s_l} \oint {\cal D} {\bf
x}(s'_l) \right] \exp \left[ - \frac{1}{4} \sum_{l=1}^N \int_0^{s_l} d
s'_l \, \dot{\bf x}^2(s'_l) \right],
\end{equation}
which we recognize as the partition function $Z$ of a grand canonical ensemble
of fluctuating closed oriented random loops, of arbitrary length and shape.
We will refer to such an ensemble as {\it loop gas}.  On account of
(\ref{det}), the exponential $\exp (-W_0)$ may alternatively be viewed as an
inverse functional determinant.  As such it is easily recognized as the
partition function of the free complex field theory,
\begin{equation}
Z = \int {\cal D} \psi^*  {\cal D} \psi \, {\rm e}^{- H},
\end{equation}
with $H$ the Hamiltonian (\ref{Hamilton}).  Thus we have proved the
equivalence of a free complex field theory and a free loop gas.

In a superconductor, the vortices have short-range interactions.  To describe
these we start again from the field theoretic side.  We shall prove that the
Hamiltonian
\begin{equation}
H = \int d^3 x \left[ |\nabla \psi|^2 +
m^2 |\psi|^2 + \lambda (|\psi|^2)^2 \right],
\end{equation}
with the additional interaction term, accounts for the steric repulsion in the
loop gas.  For this we write the complex field $\psi$ as $\psi = (\psi_1 + i
\psi_2)/\sqrt{2}$ and express the interaction term as a functional integral
over an auxiliary field $\sigma$
\begin{equation}
\exp \left[ - \frac{\lambda}{4} \int d^3 x (\psi_a^2)^2 \right] =
\int {\cal D} \sigma \exp \left[ -\int d^3 x \left( \frac{1}{\lambda}
\sigma^2 - i \psi_a^2 \sigma \right) \right],
\end{equation}
where the index $a$ runs over $1,2$.  The partition function becomes now
\begin{eqnarray}
Z &=& \int {\cal D} \psi_1 {\cal D} \psi_2 {\cal D} \sigma \exp
\left\{- \int d^3 x \left[ \frac{1}{2} (\nabla \psi_a)^2 +
\frac{m^2}{2} \psi_a^2 - i \sigma \psi_a^2 +
\frac{1}{\lambda} \sigma^2 \right] \right\} \nonumber \\ &=& \int
{\cal D} \sigma \, {\rm Det}^{-1} (- \nabla^2 + m^2 - 2i \sigma) \exp
\left( -\frac{1}{\lambda} \int d^3 x \, \sigma^2 \right) \nonumber
\\ &=& \int {\cal D} \sigma \exp \left( -\frac{1}{\lambda} \int d^3
x \, \sigma^2 \right) \sum_{N=0}^{\infty} \frac{1}{N!} \prod_{l=1}^N
\left[ \int_0^\infty \frac{d s_l}{s_l} {\rm e}^{-m^2 s_l} \oint {\cal
D} {\bf x}(s'_l) \right] \nonumber \\ & & {} \times \exp \left( -
\sum_{l=1}^N \int_0^{s_l} d s'_l \left\{ \frac{1}{4} \dot{\bf
x}^2(s'_l) -  2i \sigma [ {\bf x} (s'_l) ] \right\} \right),
\end{eqnarray}
where the last equality follows from the previous result
(\ref{explicit}).  A simple Gaussian integration yields
\begin{equation}
\int {\cal D} \sigma \exp \left\{ -\frac{1}{\lambda} \int d^3 x
\sigma^2 + 2 i \int_0^{s} d s' \, \sigma \left[ {\bf x} (s') \right] \right\}
= \exp \left\{ - \lambda \int_0^{s} d s' d s'' \, \delta \left[ {\bf
x} (s') - {\bf x} (s'') \right] \right\}.
\end{equation}
Using this in the last expression for $Z$, we obtain the real-space
representation for the partition function of a complex $|\psi|^4$ theory
\begin{eqnarray}
Z = \sum_{N=0}^{\infty} \frac{1}{N!} \prod_{l=1}^N \left[
\int_0^\infty \frac{d s_l}{s_l} {\rm e}^{-m^2 s_l} \oint {\cal D} {\bf
x}(s'_l) \right] \exp & & \left\{ - \frac{1}{4} \sum_{l=1}^N
\int_0^{s_l} d s'_l \, \dot{\bf x}^2(s'_l) \right. \nonumber \\ & & {} {}
\left. - \lambda \sum_{l,k=1}^N \int_0^{s_l} d s'_l
\int_0^{s_k} d s''_k \, \delta \left[ {\bf x} (s'_l) - {\bf x} (s''_k)
\right] \right\}.
\end{eqnarray}
This is recognized as the partition function of a loop gas with short-range
repulsion.  The field theoretic representation of an interacting loop gas by
a $|\psi|^4$ field theory will be employed in Sec.\ \ref{sec:dual} to
formulate the dual theory of a $3D$ superconductor.
\section{Dual transformation of a $3D$ lattice superconductor}
\label{sec:3DL}
In this section we set up a dual formulation of a $3D$ superconductor on a
lattice.  In contrast to the continuum case to be discussed in the next
section, the lattice model can be transformed exactly to its dual version.  We
take as starting point the partition function \cite{Peskin}
\begin{equation}
\label{LSC}
Z = \prod_{\bf x} \left[\int d {\bf A} ({\bf x}) \delta
(\overline{\nabla} \cdot {\bf A}) \right] {\rm e}^{- \frac{1}{2}
\sum_{\bf x} (\overline{\nabla} \times {\bf A})^2} \prod_{\bf x}
\left[ \int_{\pi/a}^{\pi/a} \frac{d \theta({\bf x})}{2 \pi} \right]
\exp \left\{ \beta \sum_{i,{\bf x}} \cos [ \partial_i \theta({\bf x})
- q A_i({\bf x}) ] \right\} ,
\end{equation}
where $\theta({\bf x})$ is a phase variable at site ${\bf x}$ of a
$3D$ cubic lattice, ${\bf A} ({\bf x})$ is the electromagnetic gauge
potential which is a real non-compact variable defined on the directed
links between adjacent sites, $\beta$ is the inverse temperature, and
$q$ is the electric charge.  Moreover, ${\bf x} = a x_i \bf{i}$, with
$a$ the lattice spacing, $x_i$ integers labeling the sites, and
$\bf{i}$ three orthogonal vectors spanning the lattice.  The sum
$\sum_{\bf x}$, which includes a factor $a^3$, runs over all lattice
sites, while the sum $\sum_i$ runs over all directions; $\nabla$ is a
lattice derivative with components
\begin{equation}
\partial_i f ({\bf x}) = \frac{1}{a} [ f({\bf x} + a {\bf i}) - f(
{\bf x})]; \;\;\; \overline{\partial}_i f ({\bf x}) = \frac{1}{a} [
f({\bf x}) - f({\bf x} - a {\bf i})],
\end{equation}
and the delta function $\delta (\overline{\nabla} \cdot {\bf A})$ in
(\ref{LSC}) fixes the gauge.  The lattice model (\ref{LSC}) is
appropriate for a non-compact gauge group.  We shall be working in the
Villain approximation \cite{Villain} of the model, which is obtained
by the following replacement
\begin{eqnarray}
\label{xy}
Z_{xy} &=& \prod_{\bf x} \left[ \int_{\pi/a}^{\pi/a} \frac{d
\theta({\bf x})}{2 \pi} \right] \exp \left\{ \beta \sum_{i,{\bf x}}
\cos [ \partial_i \theta({\bf x}) - q A_i({\bf x}) ] \right\}
\nonumber \\ &\rightarrow& \prod_{\bf x} \left[ \int_{\pi/a}^{\pi/a}
\frac{d \theta({\bf x})}{2 \pi} \right] \sum_{\{{\bf n} ({\bf x}) \} }
\exp \left\{ -\frac{\beta}{2} \sum_{\bf x} [ \nabla \theta({\bf x}) -
q {\bf A} ({\bf x}) - 2 \pi {\bf n} ({\bf x}) ]^2 \right\},
\end{eqnarray}
where ${\bf n} ({\bf x})$ are integers which are---like the gauge
potential ${\bf A} ({\bf x})$---defined on the directed links between
adjacent sites.  The partition function in (\ref{xy}) is given the
index $xy$ to indicate that this part of the theory is simply the
$xy$ model when the electric charge $q$ is set to zero.  We proceed by
introducing an auxiliary field ${\bf v} ({\bf x})$ via a quadratic
completion
\begin{eqnarray}
\lefteqn{ Z_{xy} = \prod_{\bf x} \left[ \int_{\pi/a}^{\pi/a}
\frac{d \theta({\bf x})}{2 \pi} \int d {\bf v} ({\bf x}) \right] }
\hspace{2.0cm} \nonumber \\ & &
\times \sum_{\{{\bf n} ({\bf x}) \} }
\exp \left\{- \frac{1}{2\beta} \sum_{\bf x} {\bf v}^2 ({\bf x}) + i
\sum_{\bf x}  {\bf v} ({\bf x}) \cdot \left[ \nabla \theta ({\bf x}) -
q {\bf A} ({\bf x}) - 2 \pi {\bf n} ({\bf x}) \right] \right\}.
\end{eqnarray}
The integration over $\theta ({\bf x})$ can then be carried out to
yield the constraint $\overline{\nabla} \cdot {\bf v} ({\bf x}) = 0$,
while the sum $\sum_{\{{\bf n} ({\bf x}) \} } \exp [- 2 \pi i {\bf v}
({\bf x}) \cdot {\bf n} ({\bf x})]$ forces the integral over the real
variable ${\bf v}$ to take on only integer values ${\bf l}$, say.  This
last observation follows from the Poisson summation formula
\begin{equation}
\label{Poisson}
\sum_n {\rm e}^{-2 \pi i v n} = \sum_l \delta (v -l).
\end{equation}
In this way the partition function $Z_{xy}$ becomes:
\begin{equation}
\label{xyspacetime}
Z_{xy} = \sum_{\{{\bf l} ({\bf x}) \} } \delta_{\overline{\nabla}
\cdot {\bf l},0 } \exp \left[ - \frac{1}{2 \beta} \sum_{\bf x} {\bf
l}^2 ({\bf x}) - i q \sum_{\bf x} {\bf l} ({\bf x}) \cdot {\bf A}
({\bf x}) \right].
\end{equation}
Whereas representation (\ref{xy}) of $Z_{xy}$ is a field theoretic
description of the lattice model, Eq.\ (\ref{xyspacetime}) describes it
in the real-space language of (charged) closed loops.  This is
analogous to the twofold description of a loop gas in the continuum
which we discussed in the preceding section.

The constraint $\overline{\nabla} \cdot {\bf l} = 0$ in
(\ref{xyspacetime}) can be explicitly solved by introducing an
auxiliary integer-valued potential ${\bf i} ({\bf x})$, via ${\bf l} =
\overline{\nabla} \times {\bf i}$, so that
\begin{equation}
\label{xydual}
Z_{xy} = \prod_{\bf x} \left[ \int d {\bf {\sf h}} ({\bf x})
\delta(\overline{\nabla} \cdot {\bf {\sf h}} ) \right] \sum_{\{{\bf m}
({\bf x}) \} } \delta_{\overline{\nabla} \cdot {\bf m},0 } \, {\rm
e}^{-2 \pi i \sum_{\bf x} {\bf m} \cdot {\bf {\sf h}} } \exp \left[ -
\frac{1}{2\beta} \sum_{\bf x} (\overline{\nabla} \times {\bf {\sf
h}})^2 - i q \sum_{\bf x} (\overline{\nabla} \times {\bf {\sf h}})
\cdot {\bf A} \right].
\end{equation}
Here, because of the presence of the factor $\exp (-2 \pi i \sum_{\bf x} {\bf
m} \cdot {\bf {\sf h}})$, the sum over the integer-valued fields ${\bf m}({\bf
x})$ ensures, on account of the Poisson formula (\ref{Poisson}), that only
integer-valued fields ${\bf {\sf h}} ({\bf x}) = {\bf i} ({\bf x})$ contribute
to the integral $\int d {\bf {\sf h}} ({\bf x})$, as required.  For $q=0$,
corresponding to the uncharged $xy$ model, Eq.\ (\ref{xydual}) gives a dual
representation in terms of vortices ${\bf m}$ coupled to a massless vector
field ${\bf {\sf h}}$.  The condition $\overline{\nabla} \cdot {\bf m} = 0$ in
(\ref{xydual}) is to assure that the vortices form closed loops, which is
physically the case.  When the integration over the field ${\bf {\sf h}}$ is
carried out, so that one is only left with vortices, one finds them
interacting via long-range forces of the Biot-Savart type.

For the lattice superconductor we have to add the electromagnetic
part to $Z_{x y}$ \cite{TS}:
\begin{eqnarray}
\label{lscdual}
Z &=& \prod_{\bf x} \left[ \int d {\bf A} ({\bf x}) \delta(
\overline{\nabla} \cdot {\bf A} ) \int d {\bf {\sf h}} ({\bf x})
\delta ( \overline{\nabla} \cdot {\bf {\sf h}}) \right] \sum_{\{{\bf
m} ({\bf x}) \} } \delta_{\overline{\nabla} \cdot {\bf m},0 }
\nonumber \\ & & {} \times \exp \left[ - \frac{1}{2} \sum_{\bf x} (
\overline{\nabla} \times {\bf A} )^2 -2 \pi i \sum_{\bf x} {\bf m}
\cdot {\bf {\sf h}} - \frac{1}{2 \beta} \sum_{\bf x}
(\overline{\nabla} \times {\bf {\sf h}})^2 - i q \sum_{\bf x}
(\overline{\nabla} \times {\bf {\sf h}}) \cdot {\bf A} \right]
\nonumber \\ &=& \prod_{\bf x} \left[ \int d {\bf {\sf h}} ({\bf x})
\delta ( \overline{\nabla} \cdot {\bf {\sf h}} ) \right] \sum_{\{{\bf
m} ({\bf x}) \} } \delta_{\overline{\nabla} \cdot {\bf m},0 } \, {\rm
e}^{-2 \pi i \sum_{\bf x} {\bf m} \cdot {\bf {\sf h}} } \exp \left\{ -
\frac{1}{2 \beta} \sum_{\bf x} \left[ (\overline{\nabla} \times {\sf
h})^2 + \beta q^2 {\bf {\sf h}}^2 \right] \right\} ,
\end{eqnarray}
where in the last step we carried out the integral over the
electromagnetic gauge potential which resulted in a mass term for the
vector field ${\bf {\sf h}}$.  The constraint $\nabla {\bf {\sf h}} =
0$ is an intrinsic part of the description of a fluctuating massive
vector field.  For a non-fluctuating field it follows automatically
from the field equation.  Physically, (\ref{lscdual}) represents a
loop gas of vortices coupled to a massive vector field.  The
massiveness of ${\bf {\sf h}}$ is in fact the only difference with the
dual description of the $xy$ model, Eq.\ (\ref{xydual}) with $q=0$.  As
a result, when integrating out this vector field, one obtains again a
Biot-Savart type of force between the vortices, but now of finite
range.

In the next section, an equivalent dual map will be carried out in the
continuum, yielding a continuum version of the lattice model (\ref{lscdual}).
The fluctuating vector field ${\bf {\sf h}}$ will be related to the physical
local induction.  With this identification in mind the constraint
$\overline{\nabla} \cdot {\bf {\sf h}}=0$ in (\ref{xydual}) can be understood
as representing the fact that the magnetic induction is divergence-free.
\section{Dual map of a $3D$ superconductor}
\label{sec:dual}
This section is the central part of the paper, in which we derive the
dual formulation of the Ginzburg-Landau model.  We shall argue that
this description is one in terms of a genuine order parameter which
involves a global rather than a local symmetry as is the case in the
Ginzburg-Landau formulation \cite{KRE}.  For this reason the dual
theory can be employed to arrive at a conventional Landau description
of the superconducting phase transition \cite{GFCM,KKR,MA}, which will
be the subject of Sec.\ \ref{sec:Landau}.

The fundamental object of the dual description is the magnetic vortex, or
Abrikosov flux tube.  Such a topological defect is either closed, infinitely
long, or---in the case of a finite system---starts and ends at the boundary of
the specimen.  A magnetic vortex can never terminate inside a superconductor.
However, for reasons that shall become clear when we proceed, we will employ a
construct that allows us to describe, at least theoretically, magnetic
vortices that do terminate inside the superconductor.  To this end we allow
the system to contain a Dirac monopole \cite{Dirac} as a test particle.
Recall that due to the Meissner effect, a magnetic field can penetrate a
superconductor only by forming quantized flux tubes.  Since also the flux
lines emanating from the monopole are squeezed into a tube, a monopole
produces precisely such a vortex.

Section \ref{sec:2D} revealed that a magnetic vortex is described by a plastic
field ${\bf B}^{\rm P}$ which appears in the theory in the combination $\nabla
\times {\bf A} - {\bf B}^{\rm P}$ with the gauge field \cite{Dirac}.  In other
words, in the presence of a test tube, the Hamiltonian becomes
\begin{equation}
\label{HP}
H^{\rm P} = \int d^3 x \left[ \frac{1}{2} (\nabla \times {\bf A} - {\bf
  B}^{\rm P})^2 + \frac{1}{2} m_A^2 A_i \left( \delta_{i j} - \frac{\partial_i
\partial_j }{\nabla^2} \right) A_j + \frac{1}{2\zeta}(\nabla \cdot
{\bf A})^2 \right],
\end{equation}
where we added a gauge-fixing term $(1/2\zeta)(\nabla \cdot {\bf A})^2$, and
given $H$ the superscript ${\rm P}$ to indicate the presence of the test
particle with its emanating tube.  We recall that we consider the
Ginzburg-Landau model (\ref{GL}) in the London limit, where the
superconducting order field is written as $\phi (x) = \exp[i\theta(x)]
w/\sqrt{2}$, with $w$ a constant.  The mass term, with $m_A = 2 e w$, is a
result of integrating out the phase field $\theta$.  In three dimensions, the
plastic field ${\bf B}^{\rm P}$ has the form \cite{Dirac,Kleinert}
\begin{equation}
\label{BP3}
B_i^{\rm P} ({\bf x}) = \frac{\pi}{e} \int_{L_{\bf z}} ds \frac{d
y_i}{d s} \delta[ {\bf x} - {\bf y} (s) ] = \frac{\pi}{e} \int_{L_{\bf
z}} d y_i \, \delta ({\bf x} - {\bf y}),
\end{equation}
which is the proper three-dimensional generalization of the $2D$ result
(\ref{BP}).  The field satisfies the equation $\nabla \cdot {\bf B}^{\rm P}
({\bf x}) = (\pi/e) \, \delta ({\bf x} - {\bf z})$ on account of Stokes'
theorem, implying that it indeed describes a monopole located at ${\bf z}$.
The line $L_{\bf z}$ starting at the point ${\bf z}$ and running to infinity
is the Dirac string accompanying the monopole.  As will be clarified
below, the location of the flux tube coincides with that of the Dirac string
(see Fig.\ \ref{fig:flux}).

From (\ref{HP}) we infer that the operator $V(L_{\bf z})$ describing
the test tube is given by
\begin{equation}
\label{dop}
V(L_{\bf z}) = {\rm e}^{- \case{1}{2} \int d^3 x \left({\bf B}^{\rm P}
\right)^2} \exp \left[ \int d^3 x \, (\nabla \times {\bf A}) \cdot {\bf
B}^{\rm P} \right].
\end{equation}
We are interested in the expectation value of this operator:
\begin{equation}
\label{VV}
\langle V(L_{\bf z}) \rangle = \int {\cal D} {\bf A} \, {\rm e}^{-
H^{\rm P}}.
\end{equation}
Since the integral over ${\bf A}$ is Gaussian, it can be evaluated by
substituting the field equation,
\begin{equation}
\label{clas}
A_i ({\bf x}) = \int d^3 y \Delta_{i j} ({\bf x} - {\bf y})
\left[\nabla \times {\bf B}^{\rm P} ({\bf y})\right]_j,
\end{equation}
back into the exponent.  The gauge-field correlation function $\Delta_{i
j}$ appearing here is
\begin{equation}
\label{gfprop}
\Delta_{i j} ({\bf x}) = \int \frac{d^3 k}{(2\pi)^3} \left(
\frac{\delta_{i j} - (k_i k_j)/{\bf k}^2}{{\bf k}^2+m_A^2} + \zeta
\frac{k_i k_j}{{\bf k}^4} \right) {\rm e}^{i {\bf k} \cdot {\bf x}}.
\end{equation}
The gauge-dependent longitudinal part of the correlation function does not
contribute to (\ref{VV}) since $\nabla \cdot (\nabla \times {\bf B}^{\rm P})
= 0$.  The expectation value is therefore independent of gauge choice.

The local induction corresponding to the classical solution (\ref{clas})
in the presence of the test tube is
\begin{equation}
\label{bclas}
\nabla \times {\bf A}({\bf x}) - {\bf B}^{\rm P}({\bf x}) =
\frac{\pi}{e} \nabla \Delta({\bf x} - {\bf z}) - m_A^2 \int d^3 y
\Delta({\bf x}- {\bf y}) {\bf B}^{\rm P} ({\bf y}),
\end{equation}
where
\begin{equation}
\label{Yuka}
\Delta ( {\bf x} ) = \int \frac{d^3 k}{( 2\pi
)^3} \frac{ {\rm e}^{i {\bf k} \cdot {\bf x}}}{{\bf k}^2+m^2_A} =
\frac{{\rm e}^{-m_A |{\bf x}|}}{4\pi |{\bf x}|}
\end{equation}
is the Yukawa potential.  The term ${\bf B}^{\rm P}$ in (\ref{bclas})
describes the Dirac string $L_{\bf z}$, the first term on the right-hand side
corresponds to the screened Coulomb force generated by the monopole.  The last
term, which is only present when $m_A \neq 0$---i.e., when there is a Meissner
effect---describes the magnetic flux tube.  A closer inspection of
(\ref{bclas}) reveals that the subtraction of the Dirac string ${\bf B}^{\rm
P}$ from the field $\nabla \times {\bf A}$ is necessary in order to obtain the
physical local induction ${\bf h}$ \cite{Dirac,Kleinert}.  Indeed, if we
calculate from the right-hand side of (\ref{bclas}) the magnetic flux through
a plane perpendicular to the Dirac string, we find
\begin{equation}
\label{Diracisright}
\int d^2 x_i \left[ \frac{\pi}{e} \partial_i \Delta({\bf x} - {\bf z})
- m_A^2 \int d^3 y \Delta({\bf x}- {\bf y}) B^{\rm P}_i ({\bf y})
\right] = - \frac{\pi}{e}
\end{equation}
that precisely one flux quantum pierces the surface in the negative direction
(see Fig.\ \ref{fig:flux}).  Here, $d^2x_i$ is an element of the surface
orthogonal to the Dirac string, and we used Gauss' law to rewrite the first
term on the left-hand side as
\begin{equation}
\frac{\pi}{e} \int d^2x_i \, \partial_i \Delta({\bf x} - {\bf z}) =
\frac{\pi}{e} \int d^3x \, \nabla^2 \Delta ({\bf x} - {\bf z}).
\end{equation}
Equation (\ref{Diracisright}) confirms Dirac's statement that the magnetic
flux emanating from a monopole must be supplied by an infinitesimally thin
string of magnetic dipoles and that in order to obtain the true local field of
a genuine point monopole, this string has to be subtracted.  While this string
is indeed completely unphysical in the normal phase, it acquires a physical
relevance in the superconducting phase \cite{Nambu} where it serves as the
core of the Abrikosov flux tube.

Substituting the field equation (\ref{clas}) back into the theory, we
obtain for the vacuum expectation value $\langle V(L_{\bf z}) \rangle$
in the London limit
\begin{equation}
\label{propstate}
\langle V( L_{\bf z} ) \rangle = {\rm e}^{\case{1}{2} \int d^3 x
\left({\bf B}^{\rm P} \right)^2} \exp \left\{ {1 \over 2} \int d^3
x d^3 y \, [\nabla \times {\bf B}^{\rm P}({\bf x})]_i \,
\Delta_{i j } ( {\bf x} - {\bf y}) \, [\nabla \times {\bf B}^{\rm
P}({\bf y})]_j \right\},
\end{equation}
or
\begin{equation}
\label{vv*}
\langle V( L_{\bf z} ) \rangle = \exp \biggl\{
- \frac{1}{2} \int d^3 x d^3 y \left[ \rho ({\bf x})\,
\Delta({\bf x} - {\bf y}) \, \rho ({\bf y}) + m_A^2 B^{\rm P}_i
({\bf x}) \, \Delta({\bf x} - {\bf y}) \, B^{\rm P}_i ({\bf y})
\right] \biggr\},
\end{equation}
where $\rho ({\bf x}) = (\pi/e) \delta ({\bf x} - {\bf z})$ is the
monopole density.  In deriving this we have omitted terms depending only
on $w$.

It should be noted that the first factor in (\ref{propstate}) diverges
\begin{equation}
\frac{1}{2} \int d^3 x \, \left( {\bf B}^{\rm P} \right)^2 =
\frac{1}{2} \left( \frac{\pi}{e}
\right)^2 \int_{L_{\bf z}} d x_i \int_{L_{\bf z}}  d  y_i
\, \delta({\bf x} - {\bf y}),
\end{equation}
representing the self-interaction of the Dirac string.  This term canceled
in (\ref{vv*}).  The first term in (\ref{vv*}) contains for ${\bf x} = {\bf
y}$ a diverging monopole self-interaction.  This divergence is irrelevant and
can be eliminated by defining a renormalized operator
\begin{equation}
\label{renop}
V_{\rm r} (L_{\bf z}) = V(L_{\bf z}) \exp\left[ \frac{\pi^2}{2 e^2}
\Delta(0) \right].
\end{equation}
The second term in (\ref{vv*}) is the most important one for our
purposes.  It represents a Biot-Savart interaction between two line
elements $d x_i$ and $d y_i$ of the magnetic vortex (see
Fig. \ref{fig:biot}).  It contains an ultraviolet singularity due to
fact that in the London limit, where the mass $|m_{\phi}|$ of the
superconducting order field is taken to be infinite, the
vortices are considered to be ideal lines.  For a finite mass the
magnetic vortices have a typical width of the order of the coherence
length $\xi =1/|m_{\phi}|$.  This mass therefore provides a natural
ultraviolet cutoff to the theory.  The last term in (\ref{vv*}) then
takes the form \cite{Nambu}
\begin{equation}
\label{moncon}
- {m_A^2 \over 2}\left(  \frac{\pi}{e}
\right)^2 \int_{L_{\bf z}} d x_i \int_{L_{\bf z}}  d y_i
\, \Delta({\bf x} - {\bf y})
= - M_V \left| L_{\bf z} \right|,
\end{equation}
with $|L_{\bf z}|$ the (infinite) length of the flux tube, and
\cite{Abrikosov}
\begin{equation}
\label{M}
M_V=\left( \frac{\pi}{e} \right)^2 \frac{m^2_A}{8\pi} \ln\left(
\frac{|m_{\phi}^2|}{m_A^2} \right) = \frac{\pi w^2}{2} \ln\left(
\frac{\lambda^2}{\xi^2} \right)
\end{equation}
being the free energy per unit length.  The combination $\lambda/\xi$
defines the dimensionless Ginzburg-Landau parameter $\kappa$, which in
the London limit is much larger than $1$.

For a monopole-antimonopole pair, (\ref{moncon}) amounts to a confining linear
potential between the monopole and antimonopole in the superconducting phase.
Let $V^*(L_{\bar {\bf z}})$ describe an antimonopole located at ${\bar {\bf
z}}$, with $L_{\bar {\bf z}}$ being a line running from infinity to ${\bar
{\bf z}}$.  Collecting all terms, we find for such a pair
\begin{equation}
\label{correlation}
\langle V_{\rm r}( L_{{\bf z}} ) V_{\rm r}^*( L_{\bar {\bf z}} )
\rangle = \exp(-M_V\left| L_{{{\bf z}}{\bar {\bf z}}}
\right|) \exp\left(\frac{\pi}{4e^2} \frac{
{\rm e}^{-m_A\left| L_{{{\bf z}}{\bar {\bf z}}} \right|} }{ \left|
L_{{{\bf z}}{\bar {\bf z}}} \right| } \right),
\end{equation}
where $L_{{{\bf z}}{\bar {\bf z}}}$ is the flux tube connecting the
monopole at ${{\bf z}}$ with the antimonopole at ${\bar {\bf z}}$, and
$|L_{{{\bf z}}{\bar {\bf z}}}|$ is its length.

We remark that the two Dirac strings may initially run to any point at
infinity.  Due to the string tension, they join on the shortest path $L_{{{\bf
z}}{\bar {\bf z}}}$ between the monopoles.  The result (\ref{correlation}) is
central to our line of arguments.  It shows that the correlation
function $\langle V_{\rm r}( L_{\bf z} ) V_{\rm r}^*( L_{\bar {\bf z}} )
\rangle$ behaves differently in the two phases \cite{KRE,Marino}.  In the
superconducting phase, where the gauge field is massive, the ``confinement''
factor dominates and the correlation function decays exponentially for
distances larger than $1/M_V$:
\begin{equation}
\label{correlator}
\langle V_{\rm r}( L_{\bf z} ) V_{\rm r}^*(
L_{\bar {\bf z}}) \rangle \stackrel{| L_{{{\bf z}}{\bar {\bf z}}} |
\rightarrow \infty }{\longrightarrow} 0.
\end{equation}
This behavior is typical for an operator in a phase without massless
excitations.  On the other hand, in the high-temperature phase, where the
gauge field is massless, the confinement factor in the correlation function
(\ref{correlation}) disappears, while the argument of the second exponential
turns into a pure Coulomb potential.  The correlation function remains,
consequently, finite for large distances:
\begin{equation}
\langle V_{\rm r}( L_{\bf z} ) V_{\rm r}^*(
L_{\bar {\bf z}} ) \rangle \stackrel{| L_{{\bf z}{\bar {\bf z}}} |
\rightarrow \infty }{\longrightarrow} 1.
\end{equation}
By the cluster property of correlation functions this implies
that the operator describing the test tube develops a vacuum
expectation value.  This signals a proliferation of magnetic vortices.
Indeed, according to (\ref{M}) the free energy $M_V$ per unit length
of a vortex vanishes at the transition point, where $w \rightarrow
0$.  It should be noted that it is the high-temperature phase and not
the superconducting phase where $V_{\rm r}(L_{\bf z})$ develops an
expectation value.

Before deriving the full-fledged dual theory let us rederive the
correlation function (\ref{correlation}) in a way that reveals some
aspects of the nature of the dual theory.  To this end we linearize
the functional integral over the gauge field by introducing an
auxiliary field ${\bf {\sf h}}$.  In the gauge $\nabla \cdot {\bf A}
= 0$, which corresponds to setting $\zeta=0$, we find
\begin{equation}
\label{Vdu}
\langle V_{\rm r}( L_{\bf z} ) V_{\rm r}^*(
L_{\bar {\bf z}}) \rangle = \int {\cal D} {\bf A} {\cal D} {\bf
{\sf h}} \exp \left[ -\frac{1}{2} \int d^3 x \, {\bf {\sf h}}^2 + i \int
d^3 x \, {\bf {\sf h}} \cdot (\nabla \times {\bf A} - {\bf B}^{\rm P}) -
\frac{m_A^2}{2} \int d^3 x \, {\bf A}^2 \right],
\end{equation}
where now $\nabla \cdot {\bf B}^{\rm P}({\bf x}) = (\pi/e) [ \delta ({\bf
x}-{\bf z})- \delta({\bf x} - {\overline {\bf z}})]$.  Also the divergence
$\nabla \cdot (\nabla \times {\bf A} - {\bf B}^{\rm P})$ is non-zero only at
the location of the artificially introduced monopoles.  In the absence of
monopoles it follows that only the transverse part of ${\bf {\sf h}}$ couples
to $\nabla \times {\bf A} - {\bf B}^{\rm P}$.  We therefore restrict the
integral over the auxiliary field ${\bf {\sf h}}$ to the transverse degrees of
freedom.  This is justified by considering the field equation for ${\bf {\sf
h}}$ following from (\ref{Vdu})
\begin{equation}
\label{phys}
{\bf {\sf h}} = i (\nabla \times {\bf A} - {\bf B}^{\rm P}) = i {\bf
h}.
\end{equation}
It tells us that apart from a factor $i$, the fluctuating field ${\bf
{\sf h}}$ may be thought of as representing the local induction
${\bf h}$, which we know to be divergence-free in the absence of
monopoles.

The integral over the vector potential is again easily carried out by
substituting the field equation for ${\bf A}$,
\begin{equation}
\label{fieldeq}
{\bf A} = \frac{i}{m_A^2}  \nabla \times {\bf {\sf h}},
\end{equation}
back into (\ref{Vdu}), with the result
\begin{equation}
\label{Vdua}
\langle V_{\rm r}( L_{\bf z} ) V_{\rm r}^*(
L_{\bar {\bf z}}) \rangle = \int {\cal D} {\bf {\sf h}} \, \delta
(\nabla \cdot {\bf {\sf h}}) \exp \left\{ -\frac{1}{2} \int d^3 x
\left[ \frac{1}{m_A^2}(\nabla \times {\bf {\sf h}})^2 + {\bf {\sf
h}}^2 \right] - i \int d^3 x \, {\bf {\sf h}} \cdot {\bf B}^{\rm P}
\right\}.
\end{equation}
We have incorporated a $\delta$ function enforcing explicitly the constraint
$\nabla \cdot {\bf {\sf h}} = 0$.  In the present formulation this constraint
is an intrinsic part of the description of a fluctuating massive vector field.
For a non-fluctuating field it is a consequence of the field equation of ${\sf
{\bf h}}$:
\begin{equation}
-\frac{1}{m_A^2} (\partial_i \partial_j - \nabla^2 \delta_{i j} ) h_j
-h_i - i B_i^{\rm P} = 0.
\end{equation}
Applying $\partial_i$ to this equation, we obtain $\nabla \cdot {\bf {\sf
h}} = 0$ provided no monopoles are present and the mass $m_A$ is
non-zero.

Expression (\ref{Vdua}) shows that the test tube described by the plastic
field ${\bf B}^{\rm P}$ couples to the fluctuating massive vector field ${\bf
{\sf h}}$, with a coupling constant given by $g = (\pi/e) m_A = 2 \pi w$ as in
two dimensions.  As $T$ approaches the critical temperature from below, $w$
goes to zero, and ${\bf {\sf h}}$ decouples from the test tube described by
${\bf B}^{\rm P}$.  After carrying out the integral over ${\bf {\sf h}}$ in
(\ref{Vdua}) we recover the result (\ref{correlation}).

The fact that the magnetic field has a finite penetration depth in
the superconducting phase is reflected by the mass term of the
${\bf {\sf h}}$ field.

It is interesting to consider the limit $m_A \rightarrow 0$ in detail,
where the massive vector field decouples from the magnetic vortex.
This limit yields the constraint $\nabla \times {\bf {\sf h}} = 0$
which can be solved by setting ${\bf {\sf h}} = \nabla \gamma$.  The
correlation function $\langle V_{\rm r}( L_{\bf z} ) V_{\rm r}^*(
L_{\bar {\bf z}}) \rangle$ then takes the simple form
\begin{equation}
\langle V_{\rm r}( L_{\bf z}) V_{\rm r}^*(
L_{\bar {\bf z}}) \rangle = \int {\cal D} \gamma \exp \left[
-\frac{1}{2} \int d^3 x (\nabla \gamma)^2 + i \int d^3 x \gamma \rho
\right],
\end{equation}
where $\rho$ is the monopole density.  In the absence of monopoles, the theory
reduces to that of a free massless mode $\gamma$ that may be thought of as
representing the magnetic scalar potential.  This follows from combining the
physical interpretation of the vector field ${\bf {\sf h}}$ (\ref{phys}) with
the equation ${\bf {\sf h}} = \nabla \gamma$.  Specifically,
\begin{equation}
\label{gammaid}
\nabla \gamma = i ( \nabla \times {\bf A} - {\bf B}^{\rm P} ).
\end{equation}
Using the definition of the monopole density, $\rho({\bf x}) = (\pi/e)
[ \delta({\bf x}-{\bf z}) - \delta({\bf x}-{\overline {\bf z}})]$, we
see that in terms of the field $\gamma$ the correlation function
reads
\begin{equation}
\label{localrep}
\langle V_{\rm r}( L_{\bf z} ) V_{\rm r}^*(
L_{\bar {\bf z}}) = \left\langle {\rm e}^{ ( \pi/e) i [\gamma ({\bf
z})- \gamma({\overline {\bf z}})] } \right\rangle.
\end{equation}
This demonstrates that the operator $V_{\rm r}( L_{\bf z} )$
describing the test tube, which was introduced in (\ref{dop}) in a
real-space formulation involving the singular plastic field ${\bf
B}^{\rm P}$ (\ref{BP3}), is now represented as an ordinary field.
Since we are in the normal conducting phase, where $V_{\rm r}$
develops a non-zero expectation value, the presence of the phase
$\gamma$ indicates that this expectation value breaks a global U(1)
symmetry, with $\gamma$ the ensuing Goldstone field.  This will be
further clarified below.

Equation (\ref{localrep}) reveals in addition that in the normal conducting
phase the Dirac string looses its physical relevance, the right-hand side
depending only on the end points ${\bf z}$ and ${\overline {\bf z}}$, not on
the line $L_{{\bf z}{\bar {\bf z}}}$.  This fact is also apparent from our
starting formula (\ref{vv*}), where the last term and therefore any reference
to $L_{\bf z}$ disappears in the limit $m_A \rightarrow 0$.  It makes no sense
to talk about magnetic vortices in this phase because they are condensed and
do not exist as physical excitations.  There is also no non-trivial topology
to assure their stability.

We are now in a position to derive the dual theory of a $3D$
superconductor.  This theory features a grand-canonical ensemble of
fluctuating closed magnetic vortices, of arbitrary shape and length,
which have a steric repulsion, i.e., a loop gas of magnetic vortices.
We know that such an ensemble can be described by a disorder field
theory, consisting of a complex $|\psi|^4$ theory.  On the other hand,
our study of a single magnetic vortex revealed that it couples with a
coupling constant $g$ to the fluctuating vector field ${\bf {\sf h}}$.
These two observations uniquely determine the dual theory in the
London limit as being given by \cite{BS,Kawai,GFCM,KKR,MA}
\begin{equation}
\label{funcZ}
Z = \int {\cal D} {\bf {\sf h}} {\cal D} \psi^{*} {\cal D} \psi
\, \delta (\nabla \cdot {\bf {\sf h}}) \exp
\left(- H_{\psi} \right)
\end{equation}
with
\begin{equation}
\label{Hpsi}
H_{\psi} = \int d^3 x \left[ \frac{1}{2 m_A^2} (\nabla \times {\bf {\sf h}})^2
+ \frac{1}{2} {\bf {\sf h}}^2 + |(\nabla -i \frac{\pi}{e} {\bf {\sf h}})
\psi|^2 + m_\psi^2 |\psi|^2 + u |\psi|^4 \right],
\end{equation}
where the $\psi$ field is minimally coupled to the vector field ${\bf {\sf
h}}$.  Equation (\ref{Hpsi}) replaces the lattice Hamiltonian (\ref{lscdual})
near the critical point.  It is a description of the superconducting state in
terms of physical variables: the field ${\bf {\sf h}}$ describes the local
induction, whereas $\psi$ accounts for the loop gas of magnetic vortices.
There are no other physical objects present in a superconductor.  The dual
theory has no local gauge symmetry because the vector field ${\bf {\sf h}}$ is
massive.  In fact, the two observations are connected.  The presence of a
local gauge symmetry in a given theory may be looked upon as reflecting a
redundancy in the description.  Since the dual theory is formulated in terms
of physical variables, there is no redundancy, and thus no local gauge
symmetry.

Although (\ref{funcZ}) was derived starting from the London limit, it is
also relevant near the phase transition.  The point is that integrating out the
size fluctuations of the scalar field $\phi$ would only generate higher-order
interaction terms and a possible change of the mass and interaction parameter
$m_\psi$ and $u$.  But these modifications do not alter the critical behavior
of the theory.

The energy $M_V$ (\ref{M}) appears in the dual theory as a one-loop
on-shell mass correction stemming from the graph depicted in Fig.\
\ref{fig:biot}, which we now interpret as a Feynman graph.  The
straight and wiggly lines represent the $\psi$ and ${\bf
A}$ field correlation functions, respectively.

A measure for the interaction strength of a massive vector field in $3D$ is
given by the dimensionless parameter equal to the square of the coupling
constant multiplied by the range of the interaction. For the dual theory this
factor is $g^2/m_A \sim m_A/e^2$, which is the inverse of the strength of the
electromagnetic gauge field ${\bf A}$ in the superconducting phase.  This is a
common feature of theories which are dual to each other.

Another notable property of the dual theory is that in the limit $e
\rightarrow 0$ it changes into a local gauge theory \cite{GFCM},
\begin{equation}
H_{\psi} \rightarrow \int d^3 x \left[ \frac{1}{2} (\nabla \times
{\bf {\sf h}})^2 + |(\nabla -i g {\bf {\sf h}}) \psi|^2 + m_\psi^2
|\psi|^2 + u |\psi|^4 \right],
\end{equation}
as can be checked by rescaling the dual field ${\bf {\sf h}}$ in the
Hamiltonian (\ref{Hpsi}).

We next investigate what happens with the dual theory when we approach the
critical temperature.  Remember that $w$ and therefore $m_A$ tends to zero in
the limit where $T$ tends to the critical temperature from below.  From the
first term in the Hamiltonian (\ref{Hpsi}) it again follows that $\nabla
\times {\bf {\sf h}} \rightarrow 0$ in this limit, so that we can write once
more ${\bf {\sf h}} = \nabla \gamma$, and (\ref{Hpsi}) becomes
\begin{equation}
\label{Hpsi'}
H_{\psi} = \int d^3 x \left[ \frac{1}{2} (\nabla \gamma)^2 + |(\nabla -i
\frac{\pi}{e} \nabla \gamma) \psi|^2 + m_\psi^2 |\psi|^2 + u |\psi|^4 \right].
\end{equation}
This equation shows that $\gamma$, representing the magnetic scalar potential,
cannot be distinguished from the phase of the disorder field.  Indeed, let
$(\pi/e) \vartheta$ be this phase.  Then, the canonical transformation
$\vartheta \rightarrow \vartheta + \gamma$ absorbs the scalar potential into
the phase of $\psi$; the first term in (\ref{Hpsi'}) decouples from the theory
and yields a trivial contribution to the partition function.  In this way, the
dual theory reduces to a pure $|\psi|^4$ theory
\begin{equation}
\label{Hpsi''}
H_{\psi} = \int d^3 x \left( |\nabla \psi|^2 + m_\psi^2 |\psi|^2 + u |\psi|^4
\right).
\end{equation}
It was already concluded that in the high-temperature phase the
magnetic vortices proliferate as indicated by the fact that $\psi$,
giving a field theoretic description of the loop gas of these
objects, develops a non-zero expectation value at the transition point.
This transition is triggered by a change in sign of $m_\psi^2$.  In
the London limit the Hamiltonian (\ref{Hpsi''}) then takes the simple
form
\begin{equation}
\label{psisim}
H_{\psi} = \int d^3 x \left[ \frac{1}{2} v^2
\left(\frac{\pi}{e}\right)^2 (\nabla \gamma)^2 \right],
\end{equation}
with $v$ the expectation value of the disorder field, $v/\sqrt{2} = \langle
|\psi| \rangle$, and where we now represented the phase of $\psi$ by $(\pi/e)
\gamma$ to bring out the fact that $\gamma$ describes the magnetic scalar
potential.  As we will demonstrate below, $v$ has the value $v = e/\pi$
\cite{KRE} of an inverse flux quantum, so that with our normalization choice
of the phase of the $\psi$ field, Eq.\ (\ref{psisim}) takes the canonical
form.

The picture of the superconducting phase transition that emerges in
the dual formulation of the Ginzburg-Landau theory is the following.
When the critical temperature is approached from below, there is a
proliferation of magnetic vortices.  We recall that in the London
limit parallel vortices repel each other, so that a single vortex
prefers to crumple.  Near $T_{\rm c}$ we then have a spaghetti of
vortices which fill the space completely at and above the transition
temperature.  Since inside the core of a vortex one has the normal
phase, the system thus becomes normal conducting.  Whereas in the
Ginzburg-Landau formulation a magnetic vortex is described by a
singular plastic field ${\bf B}^{\rm P}$, in the dual formulation it
is represented by the Noether current $j_i = \psi^*
\tensor{\partial_i} \psi -2 i g {\sf h}_i \psi^* \psi$.  This follows
from comparing the terms coupling linearly to the fluctuating ${\bf
{\sf h}}$ field.  In the normal conducting phase the field $\psi$
develops a vacuum expectation value, and thereby breaks the global
U(1) symmetry of the $|\psi|^4$ theory; $\gamma$ is the ensuing
Goldstone field.  The Noether current becomes in the London limit
${\bf j} = \nabla \gamma$, with $\gamma$ representing the massless
photon of the high-temperature phase.  It should be noted that at
$T_{\rm c}$ the fluctuating local field ${\bf {\sf h}}$ decouples from
$\psi$ because $g \rightarrow 0$.
\section{The Mixed State}
\label{sec:mixed}
In the previous section it was shown that the dual theory of the
Ginzburg-Landau model features a loop gas of magnetic vortices, i.e.,
closed random vortices of arbitrary shape and length which are
generated by fluctuations.  It is well known that magnetic vortices
can also be generated by applying an external magnetic field $H$.
Above a certain critical value $H_{{\rm c}_1}$ magnetic vortices start
to penetrate the superconductor, provided it is a type-II
superconductor.  Below this lower critical field, the Meissner effect
expels all flux lines from the system.  In thermodynamic equilibrium
the vortices in the so-called mixed state form a $2D$ triangular
lattice perpendicular to the applied field, a so-called Abrikosov flux
lattice \cite{Abrikosov} (see Fig. \ref{fig:fluxlattice}).  The $2D$
vortex density in such a lattice is given by $\rho _\otimes =
2/(\sqrt{3}a^2)$, with $a$ the lattice spacing.  The magnetic flux
$\Phi = \int d^2 x B$ through the lattice, which is given by the
number of vortices $N$ multiplied by the fundamental flux unit $\pi/e$
carried by a single flux tube, increases with the applied field.  At a
second critical value $H_{{\rm c}_2}$ the magnetic induction $B$
becomes homogeneous and saturates the applied field; no more vortices
are nucleated.  The maximum vortex density $\rho_{\otimes, {\rm max}}$
reads
\begin{equation}
\label{rmax}
\rho _{\otimes, {\rm max}} = H_{{\rm c}_2} \frac{e}{\pi } = \frac{1}{2
\pi \xi ^2},
\end{equation}
where $H_{{\rm c}_2} = 1/(2 e \xi^2)$ is the upper critical field
expressed in terms of the coherence length $\xi= 1/|m_\phi|$
\cite{saintjames}.  At this value the magnetic vortices are closely
packed and the system becomes normal conducting (see below).  The area
that can be assigned to a single vortex is $S_\otimes = 1/\rho_\otimes
= 2 \pi \xi^2$, so that $N S_\otimes$ covers the whole surface $S$
perpendicular to the applied field.

Let us study the Abrikosov flux lattice in the London limit, where the
thickness of a magnetic vortex is considered to be infinitesimal small
as compared to the penetration depth $\lambda= 1/m_A$.  In this limit,
where $|m_\phi|=1/\xi$ may be taken to be infinite, the
superconducting order field $\phi$ is frozen in, so that
there is only one type of interaction between the magnetic vortices
mediated by the gauge field ${\bf A}$.  This magnetic interaction is
repulsive for two parallel vortices.  (Outside the London limit, where
the coherence length is finite, there is a second interaction mediated
by $\phi$.  This interaction is attractive for parallel vortices.)
The Abrikosov flux lattice is easily understood in the London limit.
Due to the repulsive magnetic interaction the vortices are driven
apart, but this is called to an halt by the finiteness of the system.
The vortices then order themselves in a regular lattice---which turns
out to be a triangular lattice---so as to minimize the repulsive
interaction.

The lattice is described by the plastic field ${\bf B}^{\rm ext}$
which has only a component in say the negative third direction:
\begin{equation}
B^{\rm ext}({\bf x}) = - \sum_\alpha \frac{\pi}{e} \int d x_3 \,
\delta ({\bf x} - {\bf x}_\alpha ),
\end{equation}
where the sum is over all lattice points ${\bf x}_\alpha$, and where
we took the external field as pointing in the third direction.  In a
first approximation, the $\psi$ field in (\ref{Hpsi}) may be neglected.
The field equation for the local field ${\bf h}= (0,0,h({\bf x}))$
which we infer from (\ref{Hpsi}) augmented with the term $-i {\bf {\sf
h}} \cdot {\bf B}^{\rm ext}$ then reads
\begin{equation}
\label{euler}
-\lambda ^2 \nabla^2 h + h =  - B^{\rm ext},
\end{equation}
where we accounted for the factor $i$ between the fluctuating field
${\bf {\sf h}}$ and the physical local field ${\bf h}$.  The form
(\ref{euler}) is well-known, and can also be obtained from the
Ginzburg-Landau theory.  We consider the lattice in two different
limits.  The first limit is the one of low vortex density.  Just above
the lower critical field $H_{{\rm c}_1}$ the density is such that one
may consider the system as non-interacting.  In this case the field
equation (\ref{euler}) has the solution
\begin{equation}
\label{feld}
h({\bf x}) =  \frac{1}{2e \lambda ^2} \sum_{\alpha } K_0(\lambda
^{-1}|{\bf x}-{\bf x}_\alpha |),
\end{equation}
where $K_0$ is a modified Bessel function.  The corresponding free
energy (density) is
\begin{eqnarray}
F &=& -\frac{1}{2V} \int d^3 x \, h B^{\rm ext} \nonumber \\ &=&
\frac{\pi}{4 e^2 \lambda ^2 S} \sum_{\alpha , \beta}
K_0(\lambda^{-1}|{\bf x}_\alpha - {\bf x}_\beta |),
\end{eqnarray}
with $V$ the volume of the system, and $S$ the area perpendicular to
the applied field.  In the limit of low density one only has to
account for the self-interaction $(\alpha = \beta)$ and the
nearest-neighbor interaction, so that the free energy can be
approximated by
\begin{equation}
\label{hc1lim}
F = B H_{{\rm c}_1} \left[1+ \frac{6}{\ln (\kappa)} K_0(\lambda^{-1}a)
\right],
\end{equation}
where we used the standard result that $H_{{\rm c}_1} = \ln
(\kappa)/(4e \lambda^2)$ \cite{saintjames}.  This equation shows that
by raising the applied field an increasing part of the field energy is
used to overcome the repulsive interaction between the magnetic
vortices.  This is represented by the last term in (\ref{hc1lim}),
which not only contains a factor of $B$, like the first term, but in
addition depends on the lattice spacing $a$.  For increasing fields
$a$ decreases, implying that indeed the nearest-neighbor interaction
term becomes more important.  This observation can be nicely
illustrated by solving the thermodynamic relation $H = \partial
F/\partial B$ in terms of the magnetic induction $B$, which is related
to the vortex density via $B = (\pi /e) \rho_\otimes$ (see
Fig. \ref{fig:vortexdensity}).  For fields slightly larger than
$H_{{\rm c}_1}$, there is a rapid increase in the vortex density which
for larger fields, when the interaction between vortices becomes
important, is leveled off.

This indicates that the existence of a mixed state is due to the
repulsive magnetic interaction between magnetic vortices.  If one
takes away the interaction term in the free energy, one finds a
magnetization curve resembling that of type-I superconductors, which
do not support a mixed state.  The picture is confirmed by the
fact that the two critical field values $H_{{\rm c}_1}$ and $H_{{\rm
c}_2}$, marking the boundaries of the mixed state, become closer to
each other the smaller the value of the Ginzburg-Landau parameter
$\kappa = \lambda/\xi$ is.  With $\xi$ kept fixed, a smaller $\kappa$
value implies a smaller penetration depth which results in a stronger
shielding of the repulsive magnetic force.

There exist dirty superconductors which have a Ginzburg-Landau
parameter close to the critical value $\kappa = 1/\sqrt{2}$ separating
a type-II from a type-I superconductor.  Such a value indicates that
the screening of the two forces which are present outside the London
limit, are of the same order of magnitude.  These so-called type-II/1
materials exhibit a remarkable experimental phenomenon.  If the lower
critical field $H_{{\rm c}_1}$ is crossed from below, a whole flux
lattice jumps in instead of single vortices being produced one by one
as is the case of a deep type-II superconductor.  This is related to
the fact that for these materials the dominant force at large
distances is the attractive one mediated by the superconducting
order field \cite{KraLeu}, while at short distances the
repulsive magnetic interaction dominates (see Fig. \ref{fig:II/1}).
The lattice spacing in the flux lattice produced in such a type-II/1
superconductor is namely an equilibrium spacing resulting from the
balance between the repulsive and attractive force.  It is indicated
by $a_0$ in Fig. \ref{fig:II/1}.

Type-I superconductors, for which the two critical fields $H_{{\rm
c}_1}$ and $H_{{\rm c}_2}$ have the same value $H_{{\rm c}}$, and for
which $\kappa < 1/\sqrt{2}$, do not support a mixed state because the
screening length of the repulsive magnetic force for those materials
is smaller than the coherence length $\xi$ defining the core radius.
That is, effectively the magnetic repulsion is screened to zero
outside the vortex core, so that vortices in a type-I material
experience only an attractive force.  They will consequently coalesce,
with the result that the entire sample becomes normal.

The other limit we briefly discuss is that of high-vortex density,
where the lattice spacing $a$ is much smaller than the penetration
depth.  Because of the periodicity of the lattice, one can consider a
single unit cell and the Fourier transform becomes a series
\begin{equation}
h({\bf x}) = \sum_{\{{\bf l}\}} h({\bf k}_l)\,e^{i {\bf k}_l \cdot
{\bf x}},
\end{equation}
where ${\bf l}$ stands for the index pair ${\bf l} = (l_1,l_2)$, and
${\bf k}_l$ is the reciprocal lattice vector
\begin{equation}
{\bf k}_l = \frac{4 \pi}{\sqrt{3}a} (l_1 {\bf E}^1 + l_2 {\bf E}^2),
\end{equation}
with ${\bf E}^1=(\case{\sqrt{3}}{2},-\case{1}{2})$ and ${\bf E}^2=(0,
1)$ two vectors spanning the reciprocal lattice.  The Fourier
components of the plastic field $B^{\rm ext}({\bf x})$ are
\begin{equation}
B^{\rm ext}({\bf k}_l) = -\frac{\pi}{e} \rho_\otimes {\rm e}^{i {\bf
k}_l \cdot {\bf x}_\alpha},
\end{equation}
with ${\bf x}_\alpha$ the position of the vortex we are considering.
This leads to the free energy
\begin{equation}
\label{needsalabel}
F=\frac{1}{2}\left(\frac{\pi }{e}\right)^2 \rho_\otimes^2 \sum_{\{{\bf
l}\}} \frac{1}{1+\lambda ^2 {\bf k}_l^2}.
\end{equation}
In deriving this we used the fact that all vortices give the same
contribution, so that it suffice to consider a single vortex, say,
located at the origin ${\bf x}_{\alpha}=0$.  The term with ${\bf l} =
0$ in (\ref{needsalabel}) represents the homogeneous part.  The
remaining sum is diverging for large ${\bf l}$, but this is
artificial, stemming from the fact that we are in the London limit
where the core radius is taken to be infinitesimal small.  As before,
we take an ultraviolet cutoff of the order of the inverse coherence
length $|m_{\phi}| = 1/\xi$.  Since in the limit of large vortex
density $\lambda /a >> 1$, one can for ${\bf l} \neq 0$ neglect 1
compared to $\lambda ^2 {\bf k}_l^2$ in the denominator of
(\ref{needsalabel}).  In this way, approximating the sum by an
integral, one finds for the free energy
\begin{eqnarray}
F &=& \frac{B^2}{2}+\frac{1}{2}\left(\frac{\pi
}{e}\right)^2\rho_\otimes \int_{k_{\rm min}< |{\bf k}|< k_{\rm max}
}\frac{d^2k}{(2\pi )^2} \frac{1}{(\lambda {\bf k})^2} \nonumber \\ {}
&=& \frac{B^2}{2} + \frac{1}{2\kappa^2}
B H_{{\rm c}_2} \ln \left( \frac{k_{\rm max}}{k_{\rm min}}
\right),
\end{eqnarray}
The infrared cutoff is taken of the order of the inverse lattice
spacing $k_{\rm min} \sim 1/a$.  To be precise,
\begin{equation}
\frac{k_{\rm max}}{k_{\rm min}}=\beta \frac{a}{\xi } = \beta
\frac{\sqrt{4\pi }}{3^{1/4}}\sqrt{\frac{H_{{\rm c}_2}}{B}} \approx 1.026
\sqrt{\frac{H_{{\rm c}_2}}{B}} ,
\end{equation}
where $\beta \approx 0.381$ is a numerical factor \cite{deGennes}
which is adjusted so that the approximate calculation is in accordance
with the exact result.  One thus finds for the free energy
\begin{equation}
F \approx   \frac{B^2}{2}-\frac{1}{4\kappa ^2}BH_{{\rm c}_2} \ln
\left(\frac{B}{H_{{\rm c}_2}}\right),
\end{equation}
which in the limit $B\rightarrow H_{{\rm c}_2} $ reduces to the free
energy of the normal phase, as it should.

Intuitively, the phase transition at $H_{{\rm c}_2}$ may be pictured
as follows.  When increasing the external field, the Abrikosov flux
lattice becomes denser, and the spherical cross-section of the
normal-conducting cores will be continuously deformed into a hexagonal
one.  This picture is nicely confirmed by numerical calculations
\cite{Rammer} (see Fig. \ref{fig:Josef}).  Precisely at $H_{{\rm c}_2}$
the magnetic vortices are densely packed, and the superconducting cell
borders are squeezed to zero thickness.  In this way the mixed state
becomes a homogeneous, normal conducting state.
\section{The normal phase}
\label{sec:normal}
In Sec.\ \ref{sec:dual} we saw that the normal conducting phase is
described by a disorder theory, consisting of a $|\psi|^4$ theory in
the broken-symmetry phase.  This theory possesses again topological
defects, viz., vortices, known from the theory of superfluid $^4$He
\cite{Fetter}.  In this section a dual formulation of the
$|\psi|^4$ theory is considered in which the grand canonical ensemble
of closed vortices is described by a field theory.  The theory turns
out to be the original Ginzburg-Landau model (in the high-temperature
phase).  This is not surprising because a dual map of a dual theory
should recover the original model.

We proceed in the same manner as in Sec.\ \ref{sec:dual} and consider
a vortex ending at the point ${\bf z}$ inside the system.  To describe
such an object we would introduce in the London limit (\ref{psisim})
of the $|\psi|^4$ theory the operator
\begin{equation}
\label{vc}
W(L_{\bf z}) = {\rm e}^{-\case{1}{2} v^2 \int d^3 x \left(
\bbox{\theta}^{\rm P} \right)^2} {\rm exp}\left(\frac{\pi }{e}\,
v^2\int d^3x \, \nabla \gamma  \cdot \bbox{\theta}^{\rm P} \right),
\end{equation}
where the first factor is a singular contribution due to the core of
the vortex, cf. (\ref{dop}).  In (\ref{vc}), the factor $\pi/e$ stems
from our normalization choice of the phase of the $\psi$ field, and
the vector field $\bbox{\theta}^{\rm P}$, first introduced in the
context of superfluid $^4$He \cite{GFCM}, should be such that $\nabla
\times \bbox{\theta}^{\rm P}$ yields a delta function along the line
$L_{\bf z}$ which starts at the point ${\bf z}$ and runs to infinity
along the trajectory ${\bf y}(s)$:
\begin{equation}
\label{current}
\left( \nabla \times \bbox{\theta}^{\rm P}\right)_i ({\bf x})
= 2\pi \int_{L_{\bf z}} d s \frac{ d y_i }{d s} \delta [{\bf x} -
{\bf y}(s)],
\end{equation}
with the divergence of this equation yielding a delta function at the
point ${\bf z}$.  But this is impossible since for a regular field $
\nabla \cdot (\nabla \times \bbox{\theta}^{\rm P}) = 0$.  Hence, the
only regular $\bbox{\theta}^{\rm P}$ field that can be constructed is
one representing a closed, or infinitely long vortex $L$.  To
understand this in another way \cite{GFCM}, imagine a sphere $\Sigma$
surrounding the hypothetical endpoint of a vortex at ${\bf z}$, with a
little hole $\partial \Sigma$ where the vortex pierces the surface.
While on the one hand the loop integral $\oint_{\partial \Sigma }dx_i
\partial_i \gamma$ gives the vortex strength $2 \pi$, the surface
integral $\int_{\Sigma} d^2 x_i \epsilon_{i j k} \partial_i \partial_j
\gamma$, on the other hand, vanishes since $\gamma$ is regular on
$\Sigma$.  This contradicts Stokes' theorem which states that both
integrals should be equal and leads to the conclusion that vortices in
a $|\psi|^4$ theory can only form finite closed loops, or infinite
loops which are so to speak ``closed at infinity''.

It should be noted that the above argument does not apply to the
Ginzburg-Landau theory.  A {\em magnetic} vortex could start in a
given point by simply introducing a magnetic monopole there.  When
described with help of a gauge potential, the monopole is inevitably
accompanied by a Dirac string.  Choosing the Dirac string to pierce
the surface $\Sigma$ surrounding the monopole at $\partial \Sigma$,
one obtains $\oint_{\partial \Sigma } dx_i A_i = \int_{\Sigma} d^2 x_i
\epsilon_{i j k} \partial_j A_k$, where the right-hand side measures
the magnetic flux through the surface $\Sigma$ (see
Fig. \ref{fig:surface}).  Both sides of the equation yield the same
result, in accordance with Stokes' theorem.

For a closed loop $L$, the expectation value of $W(L)$ is given by
\begin{equation}
\label{<W>}
\langle W(L) \rangle = \int {\cal D} \gamma \exp \left[ -\frac{1}{2} v^2
\int d^3 x \left(\frac{\pi}{e} \nabla \gamma - \bbox{\theta}^{\rm
P} \right)^2 \right].
\end{equation}
The integration over $\gamma$ can be carried out by substituting the
field equation of the Goldstone field
\begin{equation}
\label{fegm}
\gamma ({\bf x})= - \frac{e}{\pi} \int d^3 y \, \Delta^0({\bf x} -
{\bf y}) \nabla \cdot \bbox{\theta}^{\rm P} ({\bf y}),
\end{equation}
where $\Delta^0$ is the correlation function (\ref{Yuka}) with
$m_A=0$.  This yields an expression
\begin{equation}
\label{ww*}
\left\langle W( L ) \right\rangle =
\exp \left\{ -{1 \over 2} v^2 \int {d^3 x }\int {d^3 y} \, [\nabla
\times \bbox{\theta}^{\rm P}({\bf x})]_i \, \Delta^0
\left( {{\bf x} - {\bf y}} \right) \, [\nabla \times
\bbox{\theta}^{\rm P}({\bf y})]_i \right\},
\end{equation}
very similar to the one obtained for a magnetic vortex in the
superconducting phase, Eq.\ (\ref{vv*}).  Using (\ref{current}), we
obtain, cf.\ (\ref{moncon})
\begin{equation}
\label{Wil}
\langle W(L) \rangle = {\rm e}^{-M_W |L|},
\end{equation}
with $|L|$ the vortex length and, cf.\ (\ref{M})
\begin{equation}
\label{MW}
M_W = \frac{\pi
v^2}{2} \ln\left(\frac{|m_{\psi}|^2}{\mu^2}\right)
\end{equation}
its energy per unit length.  In analogy with the previous calculation,
we have taken the mass $|m_{\psi}|$ of the scalar field appearing in
the theory as ultraviolet cutoff.  Since $\gamma$ is massless, $M_W$
diverges in the infrared.  This is regularized by introducing a small
mass $\mu$.

In order to find the dual theory which gives a field theoretic
description of the vortex loop gas, we rewrite the expectation value
$\langle W(L) \rangle$ via a Hubbard-Stratonovich transformation as
\begin{equation}
\label{HS}
\langle W(L) \rangle = \int {\cal D} \gamma {\cal D} {\bf b}
\exp\left\{-\int d^3 x \,  \left[\frac{1}{2} {\bf b}^2 +  i\, v
{\bf b} \cdot \left(\frac{\pi }{e}\,\nabla \gamma - \bbox{\theta}^{\rm
P}\right) \right]\right\}.
\end{equation}
The integral over the $\gamma$ fluctuations now yields the constraint
$\nabla \cdot {\bf b} = 0$, demanding ${\bf b}$ to be the rotation of
a vector field, ${\bf b} = \nabla \times {\bf A}$.  This gives
\begin{equation}
\label{sum}
\langle W (L) \rangle = \int {\cal D} {\bf A} \, {\rm exp}\left\{-\int
d^3x\, \left[ \frac{1}{2} (\nabla \times {\bf A})^2 - 2 \pi i \, v
{\bf A} \cdot {\bf J} \right] \right\},
\end{equation}
where $J_i({\bf x}) := \oint_{L} dy_i \delta ({\bf x} - {\bf y} )$
describes the closed vortex.  It is natural to interpret the
fluctuating massless gauge field ${\bf A}$ as the electromagnetic
gauge field.  This identification yields the relation we alluded to
above
\begin{equation}
\label{rel}
2 \pi \, v = 2 e
\end{equation}
between the expectation value $v$ of the disorder field $\psi$ and the
electric charge $2e$.  It is the analog of relation (\ref{g}) between
the expectation value $w$ of the superconducting order field
$\phi$ and the coupling constant $g = (\pi/e) m_A$ of the dual theory:
\begin{equation}
2\pi \, w = g.
\end{equation}

The expectation value (\ref{sum}) we now recognize as the Wilson loop.  Since
$v$ vanishes as $T$ approaches the critical temperature from above, the
coupling constant vanishes at the critical point and ${\bf A}$ decouples from
the vortex.  Precisely the same phenomenon happened with the magnetic vortex
in the superconducting phase.  Adding a gauge-fixing term to (\ref{sum}) and
carrying out the integration over the gauge field, we obtain
\begin{equation}
\langle W(L) \rangle = \langle {\rm e}^{2 i e \oint_{L} d {\bar y}_i
A_i} \rangle = \exp\left[ - 2 e^2 \int d^3 x d^3
y \, J_i({\bf x}) \Delta_{i j}^0({\bf x} - {\bf y} )J_j ({\bf y})\right],
\end{equation}
where $\Delta^0_{i j}$ is the correlation function (\ref{gfprop}) with
$m_A=0$.  Because we consider closed vortices, for which $\nabla \cdot
{\bf J}= 0$, only the first term of the correlation function
contributes, so that the result is independent of gauge choice and
is given by the previous expression (\ref{ww*}).

If we consider a loop gas of vortices, we recover, following the same
line of arguments as in Sec.\ \ref{sec:dual}, the (normal phase of the)
original Ginzburg-Landau model (\ref{GL}).  As expected, the
dual map of the dual theory gives back the original model.

Since the expectation value $v$ of the $\psi$ field vanishes when the
critical point is approached from above, it follows that also the
energy (\ref{MW}) tends to zero here.  This supports the picture that
the phase transition in a pure $|\psi|^4$ theory is associated with
the proliferation of vortices \cite{GFCM}.

The question arises now How can we physically understand a vortex loop
in the normal phase?  As is well known from $^4$He physics
\cite{Fetter,GFCM}, inside the core of a $^4$He vortex, the superfluid
order parameter vanishes, indicating that the core consists of normal
fluid.  Translated into the present context, where a finite
expectation value of the disorder $\psi$ field indicates the onset of
the normal conducting phase, we are to interpret its vanishing inside
the vortex core as the absence of the normal phase, i.e., as the
presence of the superconducting phase.  If one takes the Maxwell
equation
\begin{equation}
\nabla \times {\bf h} = 2 e {\bf j}
\end{equation}
as the defining equation of the electromagnetic current ${\bf j}$, one
has to view the purely imaginary object $2\pi i v {\bf J}$ in
(\ref{sum}) as describing an electric circuit.  (The analog between
vortices and electric currents was first pointed out by von Helmholtz.)
At first glance the presence of the factor $i$ seems strange.
In particular, the Biot-Savart law for these currents yields the
opposite sign from what is usually the case: we find two parallel
currents repelling instead of attracting each other because they carry
an imaginary charge.  (This can be checked by considering the
interaction energy $E_{\rm int}$ between two parallel vortices
\begin{equation}
E_{\rm int}= 2\pi^2 v^2 \int d x_i\int d
y_i\frac{1}{|{\bf x}-{\bf y}|},
\end{equation}
which is positive.)  But a closer inspection reveals that this has to
be the case.  Remember that an electric circuit generates a magnetic
moment
\begin{equation}
{\bf m}= e \int d^3x ({\bf x}\times {\bf j})
\end{equation}
orthogonal to the surface enclosed by the loop.  (Note that we defined
the current without a charge factor---$2e$ in our case---included,
that is why the equation for ${\bf m}$ contains a factor $\case{1}{2}
2 e = e$).  If we take two real-life circuits I and II as sketched in
Fig. \ref{fig:circuit}, where the lower laying loop I is held fixed
while the upper one is free to rotate, then the latter would settle
such that the two magnetic moments point in the same direction.  A
state where these real-life electric circuits are condensed would
inevitably be connected with a permanent magnetization.  Due to the
fact that vortex loops carry an imaginary charge, there is an
antiferromagnetic coupling rather than a ferromagnetic one between
the loops, and a vortex condensate has zero magnetization.

We note that because of the imaginary charge, the local field
generated by a vortex loop $L$ is also purely imaginary as follows
from Ampere's law,
\begin{equation}
\label{Ampere}
{\bf h}({\bf x})= i \frac{v}{2} \oint d{\bf y} \times \frac{{\bf
x}-{\bf y}}{ |{\bf x}-{\bf y}|^3} = i \frac{v}{2} \nabla \Omega ({\bf
x}),
\end{equation}
where $\Omega$ is the solid angle that the loop subtends at ${\bf x}$
(see Fig. \ref{fig:omega}).  The same result can be obtained from the
dual theory, bearing in mind that the local field, apart from a
factor $i$, can be identified with the gradient of the phase variable
$\gamma$
\begin{equation}
\label{hgamma}
{\bf h} = -i \nabla \gamma,
\end{equation}
see (\ref{gammaid}).  Rewriting the field equation (\ref{fegm}) for
$\gamma$, we find that this field can be related to the solid angle in
the following way \cite{GFCM}
\begin{equation}
\gamma({\bf x}) = \frac{e}{2\pi} \int_S d^2y_i \frac{({\bf x}-{\bf
y})_i}{|{\bf x}-{\bf y}|^3} = - \frac{e}{2\pi}\Omega,
\end{equation}
where $d^2 y_i$ is an element of the surface $S$ spanned by the loop.
Together with (\ref{hgamma}) this yields the previous result
(\ref{Ampere}).  [The magnetic moment density, or magnetization, is
represented in the dual theory by $\bbox{\theta}^{\rm P}$.  This
follows from the fact that according to (\ref{<W>}) a closed vortex
couples to the magnetic field $\nabla \gamma$ via $\bbox{\theta}^{\rm
P}$.]  The order parameter $V_{\rm r}({\bf x})$ of the normal state
essentially measures the angle $\Omega$
\begin{equation}
\label{monometer}
V_{\rm r}({\bf x}) = {\rm e}^{i \pi \gamma ({\bf x})/e} = {\rm
e}^{i\Omega ({\bf x})/2}.
\end{equation}
We recall that the operator $V({\bf x})$ was constructed by putting a
magnetic monopole at ${\bf x}$.  With this kept in mind,
Eq.\ (\ref{monometer}) becomes obvious: $\Omega/2$ is the magnetic flux
through the closed vortex $L$ emanated by the monopole.  As a last
remark we note that one can chose two topologically different surfaces
spanning the loop $L$ (see Fig. \ref{fig:monopolejail}).  Both lead,
however, to the same phase factor because $\Omega$ differs only by a
factor of $4 \pi$.
\section{Superconducting Order Parameter}
\label{sec:sop}
In the previous section it was argued that the transition to the
superconducting phase could be understood as a proliferation of (closed)
vortices of the pure $|\psi|^4$ theory.  This was concluded from the behavior
of the real-space representation (\ref{<W>}) of the single loop operator
$W(L)$, which was shown to develop an expectation value when $T$ approaches
the critical temperature from above.  The question naturally arises Is $W(L)$
related to the superconducting order field $\phi$?  To answer this question we
have to investigate how $W(L)$ is described in terms of the variables of the
Ginzburg-Landau theory.  To this end we study the object
\begin{equation}
\label{scop}
O(L_{\bf z}) = {\rm e}^{i \theta ({\bf z})}\, {\rm e}^{2 i e \int d^3
x \, {\bf A} ({\bf x}) \cdot {\bf E}^{\rm P} ({\bf x})},
\end{equation}
where $\theta$ is the phase of $\phi$, the plastic field ${\bf
E}^{\rm P}$ describes a static charge $2e$ at ${\bf z}$,
\begin{equation}
\nabla \cdot {\bf E}^{\rm P} ({\bf x}) = \delta({\bf x} - {\bf z}),
\end{equation}
and $L_{\bf z}$ is a line emanating from ${\bf z}$ and running to
infinity,
\begin{equation}
E_i^{\rm P} = \int_{L_{\bf z}} dy_i \delta ({\bf x} - {\bf y}).
\end{equation}
The second factor in (\ref{scop}) is incorporated in order
to render the operator gauge invariant. Indeed, under a gauge
transformation
\begin{equation}
{\bf A} ({\bf x}) \rightarrow {\bf A} ({\bf x}) + \nabla \Lambda ({\bf
x} ), \; \; \;  \theta ({\bf z}) \rightarrow \theta ({\bf z}) + 2e
\Lambda ({\bf z}),
\end{equation}
so that
\begin{equation}
O(L_{\bf z}) \rightarrow O(L_{\bf z}) \exp \left[ 2ie \Lambda ({\bf
z}) + 2ie \int d^3 x \, \nabla \Lambda ({\bf x}) \cdot {\bf E}^{\rm P}
({\bf x}) \right] = O(L_{\bf z}),
\end{equation}
where in the last step we performed an integration by parts.  To bring
out the gauge invariance of $O(L_{\bf z})$ more clearly we write it in
the equivalent form
\begin{equation}
O(L_{\bf z}) = \exp \left[ -i \int d^3 x ( \nabla \theta - 2 e {\bf
A}) \cdot {\bf E}^{\rm P} \right].
\end{equation}
We will be working in the low-temperature phase of the Ginzburg-Landau
model, where the gauge field is massive.  We are interested in the
expectation value
\begin{equation}
\langle O(L_{\bf z}) \rangle = \int {\cal D} {\bf A} {\cal D} \theta \,
O(L_{\bf z}) \, {\rm e}^{-H},
\end{equation}
with $H$ the Ginzburg-Landau Hamiltonian (\ref{hydroenergy}) in the
London limit.  Since both integrations are Gaussian, they are easily
carried out to yield
\begin{equation}
\label{expO}
\langle O(L_{\bf z}) \rangle = \exp \biggl\{
- \frac{1}{2} \int d^3 x d^3 y \left[ \frac{1}{w^2} n ({\bf x})\,
\Delta({\bf x} - {\bf y}) \, n ({\bf y}) + (2e)^2 E^{\rm P}_i
({\bf x}) \, \Delta({\bf x} - {\bf y}) \, E^{\rm P}_i ({\bf y})
\right] \biggr\},
\end{equation}
with $n ({\bf x}) = \delta ({\bf x} - {\bf z})$ the charge density.
This expression closely resembles the one we found for the operator
$V$ in (\ref{vv*}), describing a magnetic vortex, which can be
rewritten as
\begin{equation}
\label{expV}
\langle V( L_{\bf z} ) \rangle = \exp \biggl\{
- \frac{1}{2} \int d^3 x d^3 y \left[ \frac{1}{v^2} n ({\bf x})\,
\Delta({\bf x} - {\bf y}) \, n ({\bf y}) + g^2 {\bar
B}^{\rm P}_i ({\bf x}) \, \Delta({\bf x} - {\bf y}) \, {\bar
B}^{\rm P}_i ({\bf y}) \right] \biggr\},
\end{equation}
where ${\bar B}^{\rm P}_i$ is defined so that it contains no
factor $\pi/e$, $B^{\rm P}_i = (\pi/e) {\bar B}^{\rm P}_i$.
We see that (\ref{expO}) can be obtained from (\ref{expV}) by simply
replacing the high-temperature expectation value $v$ of the
$\psi$ field by the low-temperature expectation value $w$ of the
$\phi$ field, and by replacing the ``magnetic'' coupling $g$ by the
electric coupling $2e$.  In this sense the operators $O$ and $V$ are
dual to each other.

We continue to discuss the behavior of the expectation value
(\ref{expO}) in the two phases.  In the high-temperature phase we
argued that there can only be closed vortices.  This we achieve by
setting $n ({\bf x})$ to zero in (\ref{expO}), so that only the last
term survives.  In fact, using the relation $\pi v = e$, we
recover the right-hand side of (\ref{ww*}).  That is, the operators
$O(L)$ and $W(L)$ are the same in the high-temperature phase:
\begin{equation}
\langle O(L) \rangle = \langle W(L) \rangle = {\rm e}^{-M_W |L|},
\end{equation}
with $|L|$ the length of the vortex loop and $M_W$ given in (\ref{M}).

In the low-temperature phase the (electric) vortices are condensed and for
that reason not existing as physical excitations, only the endpoints are
physical.  The plastic field ${\bf E}^{\rm P}$ in (\ref{expO}) can then be
written as a gradient of a potential $U^{\rm P}$,
\begin{equation}
{\bf E}^{\rm P} = - \nabla U^{\rm P},
\end{equation}
with $\nabla^2 U^{\rm P} ({\bf x}) = - n({\bf x})$.  Taking a positive
charge $2e$ at ${\bf z}$ and a negative one $-2e$ at ${\bar {\bf z}}$,
we obtain for the correlation function
\begin{equation}
\langle O(L_{\bf z}) O^* (L_{\bar {\bf z}}) \rangle = \exp \left[ -
\frac{1}{2 w^2} \int d^3 x d^3 y \, n({\bf x}) \Delta^0 ({\bf x} - {\bf
y}) n ({\bf y}) \right],
\end{equation}
where now $n({\bf x}) = \delta ({\bf x} - {\bf z}) - \delta ({\bf x} -
{\bar {\bf z}} )$, and $\Delta^0 ({\bf x} - {\bf y})$ is the {\it
massless} scalar correlation function.  For ${\bf x} = {\bf y}$ we
have again a diverging self-interaction which is irrelevant and can
be eliminated by defining a renormalized operator $O_{\rm r}$ in the
same way as we did before in (\ref{renop}).  We then find
\begin{equation}
\langle O_{\rm r}({\bf z}) O_{\rm r}^* ({\bar {\bf z}}) \rangle =
\exp \left( \frac{1}{4 \pi w^2} \frac{1}{|L_{ {\bf z} {\bar {\bf
z}}}|} \right).
\end{equation}
This low-temperature expression is completely analogous to the one in
the high-temperature phase for the correlation function $\left\langle
V_{\rm r}({\bf z}) V_{\rm r}^*({\bar {\bf z}}) \right\rangle$,
Eq.\ (\ref{correlation}).  Using the relation $e = \pi v$, we can write
the latter as
\begin{equation}
\left\langle V_{\rm r}({\bf z}) V_{\rm r}^*({\bar
{\bf z}}) \right\rangle = \exp \left(\frac{1}{4 \pi v^2} \frac{1}{|
L_{{\bf z} {\bar {\bf z}}} |} \right).
\end{equation}
For large separation $\langle O_{\rm r}({\bf z}) O_{\rm r}^* ({\bar
{\bf z}}) \rangle \rightarrow 1$, implying that $O_{\rm r}({\bf z})$
develops an expectation value in the superconducting phase.  That is,
$O_{\rm r}({\bf z})$ is the superconducting order parameter.  Being
gauge invariant this operator makes no statement about the local U(1)
symmetry.  Referring back to the first representation (\ref{scop}) of
the superconducting order parameter, we find that a non-zero
expectation value indicates that the {\it global} U(1) symmetry
parameterized by a constant transformation parameter $\Lambda_0$ is
spontaneously broken.

In closing this section, we remark that the superconducting order
parameter can also be represented in the dual theory.  The result is
that ${\bf E}^{\rm P}$ appears in (\ref{Hpsi}) in the combination
$\nabla \times {\bf {\sf h}} - 2e {\bf E}^{\rm P}$ with the
fluctuating ${\bf {\sf h}}$ field.  To derive this result it is
prudent not to proceed in the manner we exploited before to obtain the
dual theory (\ref{Hpsi}) and linearize the gauge-field fluctuations,
but instead linearize the $\theta$ fluctuations.  The result can also
be inferred using a duality argument, remembering that the disorder
parameter was incorporated in the original theory by the combination
$\nabla \times {\bf A} - {\bf B}^{\rm P}$, see Eq.\ (\ref{HP}).
\section{Disorder field theory for the superconducting phase transition}
\label{sec:Landau}
We have argued that the dual description of the Ginzburg-Landau model is one
in terms of physical variables, and that it possesses no local gauge symmetry.
A disorder field $\psi$ was identified which vanishes in the superconducting
phase, and which develops an expectation value in the normal conducting phase,
thereby breaking a global U(1) symmetry.  We recall that central to Landau's
theory of continuous phase transitions is the presence of an order parameter
which signals through its vacuum expectation value whether or not a certain
symmetry is broken.  This is precisely what the disorder field $\psi$ does.
On that ground $\psi$ is ideally suited to formulate a Landau type of
description of the superconducting phase transition \cite{KRE,KKR}.  To
understand why the Ginzburg-Landau theory itself is not well suited to do
this, we note that it has a {\it local} gauge symmetry.  According to
Elitzur's theorem \cite{Elitzur} such a symmetry can never be broken, so that
for a local symmetry no order parameter exists in the sense of Landau.  This
may be one of the reasons for the fact that no infrared stable fixed point was
found within the Ginzburg-Landau formulation of the superconducting
phase transition \cite{HLM,KS}, although it is generally accepted that the
transition is of second order in the type-II regime, and thus should possess
such a point.

Below the transition temperature we saw that the dual theory consists
of a $|\psi|^4$ theory coupled to a massive vector field.  Above
$T_{\rm c}$ this field decouples, and the disorder field $\psi$
simultaneously develops a vacuum expectation value.  Despite the
apparent difference in the description of the low- and
high-temperature phase, it is readily argued that at the mean-field
level the critical behavior is governed by a simple $|\psi|^4$ theory.
This can be seen by integrating out the massive vector field in the
low-temperature phase.  Apart from irrelevant terms, this leads to
only a change in the coefficients of the $|\psi|^4$ theory, no
additional relevant terms such as $|\psi|^3$ are generated.
Explicitly,
\begin{equation}
H_{\psi,{\rm eff}} = \int d^3 x \left[ |\nabla \psi|^2 +
\left(m_\psi^2 - g^2 \frac{m_A}{2 \pi} \right) |\psi|^2 +
\left(u -  \frac{g^4}{4\pi m_A} \right) |\psi|^4 \right].
\end{equation}
In deriving this effective Hamiltonian we used dimensional regularization;
(irrelevant) higher-order terms were omitted.  We note that all contributions
stemming from the vector field ${\bf {\sf h}}$ vanish in the limit $T$
approaching $T_{\rm c}$ from below, so that $H_{\psi,{\rm eff}}$ reduces to
(\ref{Hpsi''}) in this limit.  It is well known that a $|\psi|^4$ theory with a
positive coupling has a non-trivial infrared stable fixed point and undergoes a
second-order phase transition.  We therefore conclude that at the mean-field
level also the superconducting phase transition is of second order if the
system is sufficiently deep in the type-II regime.  The same
conclusion can be reached starting from the Ginzburg-Landau formulation.

Below we will apply renormalization group theory to see if this
conclusion holds also beyond mean-field theory.  Halperin, Lubensky,
and Ma \cite{HLM} performed this study within the Ginzburg-Landau
theory.  Using an $\epsilon$ expansion, they showed that at the
one-loop level the theory no longer possesses an infrared stable fixed
point.  They interpreted this as indicating that the transition is of
first order.  This conclusion was in accordance with results obtained
by Coleman and Weinberg \cite{CW} who studied the electrodynamics of
{\it massless} scalar mesons in four dimensions and discovered that at
the one-loop level the photon acquires a mass.  A study of the
effective action shows a precocious onset of the Higgs mechanism with
a sudden appearance of a finite photon mass.  This is typical for a
first-order transition.  Only by artificially enlarging the number of
components of the complex scalar field $\phi$ appearing in the
Ginzburg-Landau model did Halperin, Lubensky, and Ma find an infrared
stable fixed point, provided this number is taken to be larger as 183.
However, the corresponding critical exponent $\eta$, which determines
the anomalous dimension of $\phi$, depends on the gauge-fixing choice
and is therefore unphysical.  This should not come as a surprise since
$\phi$ itself is not gauge invariant, and therefore not physical.

We shall carry out the renormalization group theory within the dual
formulation which, being casted in terms of physical fields, does not
suffer from the flaws of local gauge invariance \cite{KKS}.  The dual
theory (\ref{Hpsi}) involving a massive vector field is perturbatively
renormalizable in four dimensions $(D=4)$ \cite{Collins}, so that
usual perturbation theory can be applied to calculate the critical
exponents.  However, the derivation of the dual theory (\ref{Hpsi})
from the Ginzburg-Landau model hinged on the fact that the number of
space dimensions is three, for which the dual object $\epsilon _{i j
k } \partial_j A_k$ is a vector.  In other words, the dual theory
describes the superconducting phase transition only in three
dimensions.  For this reason, we carry out the renormalization group
theory in fixed ($D=3$) dimension, and not in $D=4-\epsilon$
dimensions as is often done.  The fixed-dimension approach to critical
phenomena was introduced by Parisi, who applied it to a pure
$|\psi|^4$ theory \cite{Plect,Parisi}.  The method makes explicitly
use of the fact that near the critical point the system has only one
relevant length scale, viz., the correlation length which diverges at
this point.  This length is used to convert dimensionful coupling
constants into dimensionless ones.

In the present setting the relevant scale is the (renormalized)
inverse mass $m_\psi^{-1}$. (The bare mass vanishes as $m_{\psi,0}^2
\sim T_{\rm c}-T$ at the critical temperature $T_{\rm c}$).  We know
from the Ginzburg-Landau theory that the bare penetration depth also
diverges at $T_{\rm c}$, viz., $\lambda_0 \sim (T_{\rm c}-T)^{-1/2}$.
However, the renormalized length $\lambda$ should not
constitute an independent diverging length scale.  We will see below
that this is indeed the case.  As usual, the critical exponents are
computed in the symmetric phase of the model, which in the present
context corresponds to the superconducting phase.

We write the bare Hamiltonian (\ref{Hpsi}) as a sum of the renormalized
Hamiltonian and counterterms $\delta H$
\begin{eqnarray}
\label{counter}
\delta H = \int d^3x & & \left[ (Z_\psi - 1) |(\nabla -i g \bbox{\sf h})
\psi|^2 + (Z_\psi m_{\psi,0}^2 - m_\psi^2) |\psi|^2 + u (Z_u-1)
|\psi_0|^4   \right. \nonumber \\ & & \left.
+ \frac{1}{2}(Z_{{\sf h}} - 1) (\nabla \times
{\bf {\sf h}})^2 + \frac{1}{2} (Z_{{\sf h}} m_{A,0}^2- m_A^2)
{\bf {\sf h}}^2 \right].
\end{eqnarray}
(All quantities appearing in (\ref{Hpsi}) should have been given an
index 0 to indicate that they refer to bare quantities.  For
convenience we rescaled the $\bbox{\sf h}$ field by a factor $m_A$:
$\bbox{\sf h} \rightarrow m_A \bbox{\sf h}$.)  The renormalized
objects are related to the bare ones via
\begin{equation}
\label{r-b}
{\sf h}_{i} = Z_{{\sf h}}^{-1/2} {\sf h}_{0,i} , \;\;\;
g = Z_g^{-1} Z_\psi Z_{{\sf h}}^{1/2} g_0, \;\;\;
\psi = Z_{\psi}^{-1/2} \psi_0 , \;\;\;  u = Z_{u}^{-1} Z_{\psi}^2
\, u_0.
\end{equation}
It is straightforward to calculate the one-loop diagrams.  The correlation
functions can be read off from the Hamiltonian (\ref{Hpsi}).  With
a wiggly line denoting the correlation function of the vector
field, and a straight line denoting the one of the $\psi$ field, it follows
that
\begin{mathletters}
\begin{eqnarray}
\epsfbox{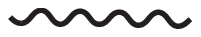} &=& \frac{1}{{\bf k}^2 + m_A^2} \left( \delta_{ij} -
\frac{k_i k_j}{{\bf k}^2} \right) \label{hcorr} \\
& & \nonumber \\
\epsfbox{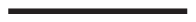} &=& \frac{1}{{\bf k}^2 + m_\psi^2}.
\end{eqnarray}
\end{mathletters}
The correlation function (\ref{hcorr}) reflects the constraint $\nabla \cdot
\bbox{\sf h}= 0$ which we imposed upon the fluctuating field $\bbox{\sf h}$.
The correlation function has been obtained by representing the $\delta$
function $\delta (\nabla \cdot \bbox{\sf h})$ as $\exp[-(\nabla \cdot
(\bbox{\sf h})^2/(2 \zeta)]$ with $\zeta$ taken to zero at the end.  We
find for the diagrams depicted below
\begin{mathletters}
\begin{eqnarray}
\raisebox{-0.25cm}{\epsfbox{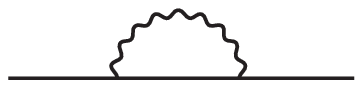}}\hspace{2cm}
&=& \frac{2}{3\pi} \frac{g^2}{m_\psi+ m_A} {\bf k}^2 \label{dia1} \\
& & \nonumber \\
\raisebox{-0.5cm}{\epsfbox{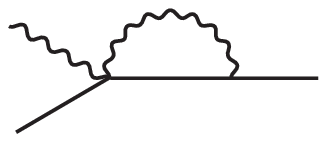}}\hspace{2.2cm}
&=& i \frac{2}{3\pi} \frac{g^3}{m_\psi+ m_A} k_i
\label{dia2} \\ & & \nonumber \\
\raisebox{-0.62cm}{\epsfbox{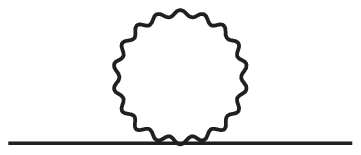}}
\, \, + \, \, \raisebox{-0.62cm}{\epsfbox{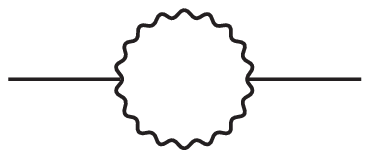}}
 &=& -\frac{1}{24 \pi }
\frac{g^2}{m_\psi} {\bf k}^2 \left( \delta_{i j} - \frac{k_i k_j}{{\bf
k}^2} \right). \label{dia3} \\ & & \nonumber
\end{eqnarray}
\end{mathletters}
We considered these particular diagrams to illustrate the following
observations.  The first diagram yields a $Z_\psi$ factor given by
\begin{equation}
Z_\psi = 1 + \frac{2}{3\pi} \frac{g^2}{m_\psi+ m_A}.
\end{equation}
From the second diagram we extract the factor $Z_g$,
\begin{equation}
Z_g = 1 + \frac{2}{3\pi} \frac{g^2}{m_\psi+ m_A},
\end{equation}
which turns out to be equal to $Z_\psi$.  To this order, the renormalized
coupling constant $g$ is thus related to the bare one simply via
\begin{equation}
g = Z_{\sf h}^{1/2} g_0.
\end{equation}
This means that the minimal coupling to the vector field is preserved
at the one-loop level.  In fact, it is preserved to any order in the
loop expansion.  The reason is that the Ward identity which guarantees
that the minimal coupling is always maintained in the case of a massless
vector field, also operates in the massive case \cite{Collins}.

The diagrams (\ref{dia3}) show that the one-loop contributions to the
self-energy of the vector field is transverse.  This too remains true
to all orders in perturbation theory thanks to the Ward identity.  The
mass term of the vector field is consequently not renormalized and
does not need a counterterm.  That is, $m_{A} = Z_{{\sf h}}^{1/2}
m_{A,0}$, so that the last term in (\ref{counter}) vanishes.  It
follows that the critical exponent $\gamma_{{\sf h}}$ is
unaffected by the fluctuations, and retains its mean-field value
$\gamma_{{\sf h}} = 1$.

Incidently, the electric charge does not renormalize in the dual theory since
$g = (\pi/e) m_A$, and both $g$ and $m_A$ renormalize in the same manner.

We now come to an important observation related to the fact that in
the dual theory the inverse penetration depth $m_A$ plays the role of
a mass as well as of a coupling constant because $g = (\pi/e) m_A$,
with $e$ a constant.  The standard definition of the critical exponent
$\nu$ which determines how the correlation length $m_\psi^{-1}$
diverges when the temperature approaches $T_{\rm c}$: $m_\psi^{-1}
\sim (T_{\rm c} - T)^{-\nu}$, is
\begin{equation}
\label{nupsi}
\frac{1}{\nu} = \frac{\partial \ln(m_{\psi,0}^2)}{\partial \ln
(m_\psi)}.
\end{equation}
In our case this can be rewritten as follows
\begin{equation}
\label{rew}
\frac{\partial  m_{A,0}^2}{\partial \ln (m_\psi)} =
\frac{m_{A,0}^2}{\nu}
\end{equation}
because $m_{\psi,0}^2 \sim m_{A,0}^2$ near $T_{\rm c}$.  The
$\beta$ function is defined by the equation
\begin{equation}
\beta (\hat{g}^2) := m_\psi \frac{\partial }{\partial  m_\psi} \left.
\frac{g^2}{m_\psi} \right|_{u_0,g_0},
\end{equation}
with the properly scaled coupling constant $\hat{g}^2 := g^2/m_\psi$.
By virtue of (\ref{rew}), this can be cast in the form
\begin{equation}
\label{beta}
\beta (\hat{g}^2) = \hat{g}^2 \left( -1 + \frac{1}{\nu} +
\gamma_{{\sf h}}(\hat{g}^2,\hat{u}) \,  \frac{\partial
\ln(m_A)}{\partial \ln (m_\psi)} \right),
\end{equation}
where $\gamma_{{\sf h}}(\hat{g}^2,\hat{u})$, with $\hat{u}:=u/m_\psi$,
is the function
\begin{equation}
\gamma_{{\sf h}}(\hat{g}^2,\hat{u}) := m_A \frac{\partial }{\partial m_A}
\ln(Z_{{\sf h}})|_{u_0,g_0}
\end{equation}
which yields the critical exponent $\eta_{{\sf h}}$ when evaluated at
the critical point.  Without the explicit mass dependence, the
coefficient of the $\hat{g}^2$ term in the $\beta
(\hat{g}^2)$ function would be $-1$, implying that the origin
$\hat{g}=0$ is an ultraviolet stable fixed point.  In (\ref{beta}),
however, the coefficient is $-1+1/\nu$ which is positive if $\nu < 1$.
In this case, the origin $\hat{g}^2=0$ becomes infrared stable and the
coupled theory reduces to a pure $|\psi|^4$ theory.  The best estimate
for $\nu$ available from summed perturbation theory at fixed $D=3$
\cite{Z-J} gives $\nu \approx .6695$, which is smaller than one.
Hence, the trivial fixed point $\hat{g}^{*^{\scriptstyle 2}} = 0$ is
infrared stable (see Fig. \ref{fig:stable}).  This situation differs
dramatically from that in the Ginzburg-Landau theory where the
coupling $e$ to the electromagnetic gauge field has an infrared stable
fixed point away from the origin, see Fig. \ref{fig:unstable}, and the
corresponding value $\hat{e}^{*^{\scriptstyle 2}}$ is too large to
allow the coupled system to develop an infrared stable fixed point.

To recapitulate, the dual theory of the superconducting phase
transition possesses an infrared stable fixed point given by
$\hat{g}^{*^{\scriptstyle 2}} = 0$ and $\hat{u}^* = \hat{u}^*_{\rm
WF}$, where $\hat{u}^*_{\rm WF}$ is the Wilson-Fisher fixed point of a
pure $|\psi|^4$ theory with reversed temperature axis.  The critical
exponents of the $\psi$ field are the ones of a superfluid.  The
critical exponents pertaining to the $\bbox{\sf h}$ field, which
physically represents the local induction, have their mean-field
values.  In particular, $\nu_{\sf h} = 1/2$. This exponent reveals
that the magnetic penetration depth diverges near $T_{\rm c}$ as
$(T_{\rm c}-T)^{-1/2}$, meaning that inside the critical region the
empirical formula $\lambda \sim [1-(T/T_{\rm c})^4]^{-1/2}$ found
outside this region remains unchanged.

A last point of interest is the Gaussian fixed point, corresponding to
${\hat g}^{*^{\scriptstyle 2}} = 0, \hat{u}^* = 0$.  This fixed point
is infrared stable in the $\hat{g}^2$ direction, and unstable in the
${\hat u}$ direction.  It describes a loop gas of free vortices.  At
this point the phase transition changes from second to first order,
i.e., it is a tricritical point, the existence and location of which
was first established in Ref.\ \cite{tricritical}.  At the level we are
working, the critical exponents characterizing this point are Gaussian.
A $|\psi|^6$ term which should now be included will generate
logarithmic corrections.
\acknowledgments
We thank A. Kovner and J. Rammer for useful discussions.
\appendix
\section*{}
Here we give a lattice version of the arguments presented
in Sec.\ \ref{sec:twofold}, showing the equivalence of a complex field
theory and a loop gas \cite{GFCM,AID}.  We start with the free
Hamiltonian (\ref{Hamilton}) defined on a hypercubic lattice in $D$
space dimensions.  (We use the same lattice notation as in Sec.
\ref{sec:3DL}).  The corresponding energy expression we write as
\begin{equation}
\label{lenergy}
E = \sum_{{\bf x}, {\bf y}} a^{D-2} \psi^* ({\bf x}) \Lambda({\bf x}, {\bf
y}) \psi ({\bf y}),
\end{equation}
where
\begin{equation}
\Lambda({\bf x}, {\bf y}) = (2D + m^2 a^2) \delta_{{\bf x}, {\bf y}} -
\sum_i \left( \delta_{{\bf x}, {\bf y} + a {\bf i} } +
\delta_{{\bf x}, {\bf y} - a {\bf i} } \right),
\end{equation}
or in matrix notation
\begin{equation}
\Lambda = (2D + m^2 a^2) I -S.
\end{equation}
Here, $I$ is the identity operator and $S$ is the so-called step
operator.  The matrix element $S({\bf x}, {\bf y})$ is $1$ if the two
lattice sites ${\bf x}$ and ${\bf y}$ are neighbors, and zero
otherwise.   From (\ref{lenergy}) we  immediately read off the
lattice correlation function
\begin{equation}
\label{lgreen}
G( {\bf x}) = a^{2-D} \Lambda^{-1}(0,{\bf x}).
\end{equation}
The operator $\Lambda^{-1}$ we expand in a von Neumann series
\begin{equation}
\label{Lambda}
\Lambda^{-1} = \frac{\sigma}{2D} \sum_{N=0}^{\infty} \left( \frac{ S
\sigma }{2 D} \right)^N,
\end{equation}
where we introduced the abbreviation
\begin{equation}
\sigma^{-1} = 1 + \frac{m^2 a^2}{2D}.
\end{equation}
From the definition of the step operator $S$ it follows that $S^N({\bf
x}, {\bf y})$ is equal to the number of paths which go from site ${\bf
x}$ to site ${\bf y}$ in $N$ steps.  By virtue of this we can write
\begin{equation}
\Lambda^{-1}(0,{\bf x}) = \frac{\sigma}{2 D} \sum_{N=0}^{\infty}
P_N ({\bf x}) \sigma^N,
\end{equation}
with $P_N ({\bf x})$ the probability defined by the number of paths
that go from $0$ to ${\bf x}$ in $N$ steps divided by the ones
that just start in $0$ and contain $N$ steps.  This last number
is equal to $(2D)^N$.  The probability $P_N ({\bf x})$ plays a central
role in the theory of random walks.  A little thought reveals that it
satisfies the recurrence relation
\begin{equation}
\label{rec}
P_{N+1} ({\bf x}) = \frac{1}{2D} \sum_i P_N ({\bf x}+ a {\bf i}),
\end{equation}
with the initial condition
\begin{equation}
\label{init}
P_0 ({\bf x}) = \delta_{0,{\bf x}}.
\end{equation}
To solve (\ref{rec}) we introduce the Fourier transform:
\begin{equation}
P_N ({\bf x}) = a^D \int_{-\pi/a}^{\pi/a} \frac{d^D k}{(2 \pi)^D}
{\rm e}^{i {\bf k} \cdot {\bf x}} P_N ({\bf k}).
\end{equation}
The recurrence relation then yields for $P_N ({\bf k})$
\begin{equation}
P_{N+1} ({\bf k}) = \frac{1}{D} \sum_i \cos(k_i a) P_N ({\bf k}),
\end{equation}
with $P_0 ({\bf k}) = 1$ as follows from the initial condition
(\ref{init}).  In this way we obtain as solution
\begin{equation}
\label{prob}
P_N ({\bf x}) = a^D \int_{-\pi/a}^{\pi/a} \frac{d^D k}{(2 \pi)^D}
{\rm e}^{i {\bf k} \cdot {\bf x}} \left[ \frac{1}{D} \sum_i \cos (k_i
a) \right]^N.
\end{equation}

We are now in a position to take the continuum limit ($a \rightarrow
0$).  Let us first concentrate on the probability (\ref{prob}) and
assume that the length $s$ of a path on a lattice is measured in
some---as yet arbitrary---unit $b$.  That is, a path of $N$ steps has
length $s = N b$.  Since in the limit $a \rightarrow 0$
\begin{equation}
\left[ \frac{1}{D} \sum_i \cos (k_i a) \right]^N \rightarrow \left(1 -
\frac{a^2}{2D} {\bf k}^2 \right)^{s/b} \rightarrow \exp \left(-s
\frac{a^2}{2 D  b} {\bf k}^2 \right),
\end{equation}
we have to take the ratio $a^2/b$ fixed in order to obtain a
non-trivial continuum limit; we will take $a^2/b = 2 D$.  In other
words, $b \sim a^2$ in the scaling limit.  In this limit the probability
density $p({\bf x},s) := \lim_{a \rightarrow 0} P_N ({\bf x})/a^D$
becomes
\begin{equation}
\label{pdensity}
p({\bf x},s) = \int \frac{d^D k}{(2 \pi)^D} {\rm e}^{i {\bf k} \cdot
{\bf x}} {\rm e}^{- s {\bf k}^2 },
\end{equation}
so that the correlation function (\ref{lgreen}), with $\Lambda^{-1}$
given by (\ref{Lambda}), can be written as
\begin{eqnarray}
G({\bf x}) &=& \lim_{a \rightarrow 0} a^{2-D} \Lambda^{-1}(0,{\bf x})
= \lim_{a \rightarrow 0} \frac{a^2 \sigma}{2 D} \sum_{N=0}^{\infty}
\frac{1}{a^D} P_N ({\bf x}) \left(1 + \frac{m^2 a^2}{2D} \right)^{-s/b}
\nonumber \\  &=& \int_0^{\infty} {\rm e}^{-s m^2} p({\bf x},s),
\end{eqnarray}
where we used the fact that $\sigma \rightarrow 1$ and
\begin{equation}
\frac{a^2}{2D} \sum_{N=0}^{\infty} \rightarrow \int_0^{\infty} ds
\end{equation}
when $a \rightarrow 0$.  But with $p({\bf x},s)$ given in
(\ref{pdensity}), this is precisely the Schwinger representation
(\ref{green}) of the correlation function.

A similar derivation can be given for the partition function.  To this
end we consider the probability $P^{\rm c}_N$ defined by the number of
closed oriented paths containing $N$ steps divided by $(2 D)^N$.  It
is related to the probability $P_N({\bf x})$ previously introduced via
\begin{equation}
P_N(0) = N P_N^{\rm c},
\end{equation}
where the factor N arises from the fact that one can start traversing
a given oriented loop at any of the $N$ lattice sites visited by the
loop.  It then easily follows that
\begin{equation}
\sum_{N=0}^{\infty} \frac{P_N^{\rm c}}{a^D} \sigma^N
\rightarrow \int_0^{\infty} \frac{ds}{s}  {\rm
e}^{-s m^2} \int \frac{d^D k}{(2 \pi)^D} {\rm e}^{-s {\bf k}^2},
\end{equation}
which is the right-hand side of the first equation in (\ref{oint}).
This leads to the same representation of the partition function we
found before with help of Schwinger's proper-time representation.  The
steric repulsion can also be included on the lattice, but this will
not be done here.
\begin{figure}
\caption{ Magnetic field (thin lines) of a positively charged
monopole.  The magnetic flux is provided by the Dirac string (thick line)
located along the negative $z$ axis.
\label{fig:flux} }
\end{figure}

\begin{figure}
\caption{Biot-Savart interaction (wiggly line) between two line elements
$d x_i$ and $d y_i$ of a magnetic vortex (straight line).
\label{fig:biot} }
\end{figure}

\begin{figure}
\caption{Triangular Abrikosov flux lattice.  The two vectors
${\bf v}_1$ and ${\bf v}_2$ span the unit cell, containing one flux
tube.  \label{fig:fluxlattice} }
\end{figure}

\begin{figure}
\caption{Magnetic induction $B$ versus applied magnetic field $H$ in
arbitrary units for a type-II superconductor.  \label{fig:vortexdensity}}
\end{figure}

\begin{figure}
\caption{Schematic representation of the interaction potential $V(r)$
between two parallel magnetic vortices for a type-II/1 superconductor.
\label{fig:II/1}}
\end{figure}

\begin{figure}
\caption{Contour plots of the local induction $h$ in an Abrikosov
flux lattice for a type-II superconductor with $\kappa=1.3$.  The
first plot is at $B/H_{{\rm c}_2} = 0.1$ and the second is at
$B/H_{{\rm c}_2} = 0.9$ (after
Ref.\ \protect\cite{Rammer}). \label{fig:Josef}}
\end{figure}

\begin{figure}
\caption{A magnetic monopole with its string piercing the surface
$\Sigma$ at $\partial \Sigma$. \label{fig:surface} }
\end{figure}

\begin{figure}
\caption{Two magnetic moments ${\bf m}_I$ and ${\bf m}_{II}$
interacting with each other.  The first one is kept fixed, while the
second is free to rotate.  \label{fig:circuit} }
\end{figure}

\begin{figure}
\caption{Solid angle $\Omega$ that the vortex loop ${\bf J}$ subtends
at the point ${\bf x}$.  \label{fig:omega} }
\end{figure}

\begin{figure}
\caption{Two different surfaces spanning the loop $L$. The flux
through the surfaces differ by a factor $4 \pi$.
\label{fig:monopolejail} }
\end{figure}

\begin{figure}
\caption{The $\beta$ function for the dual coupling constant (squared)
$\hat{g}^2 = g^2/m_\psi$, showing that the origin is an infrared
stable fixed point.
\label{fig:stable}}
\end{figure}

\begin{figure}
\caption{The $\beta$ function for the electric charge (squared)
$\hat{e} = g/m_\psi$ as obtained from the Ginzburg-Landau theory,
showing that the origin is an infrared unstable fixed point.
\label{fig:unstable}}
\end{figure}
\newpage

\begin{center}
\makebox{\epsfbox{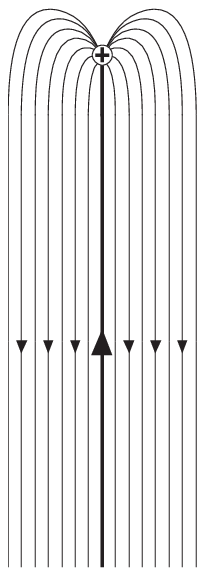}} \\[1cm]
FIG. 1\\[2cm]
\makebox{\epsfbox{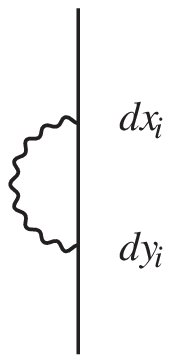}} \\[1cm]
FIG. 2\\[2cm]
\makebox{\epsfbox{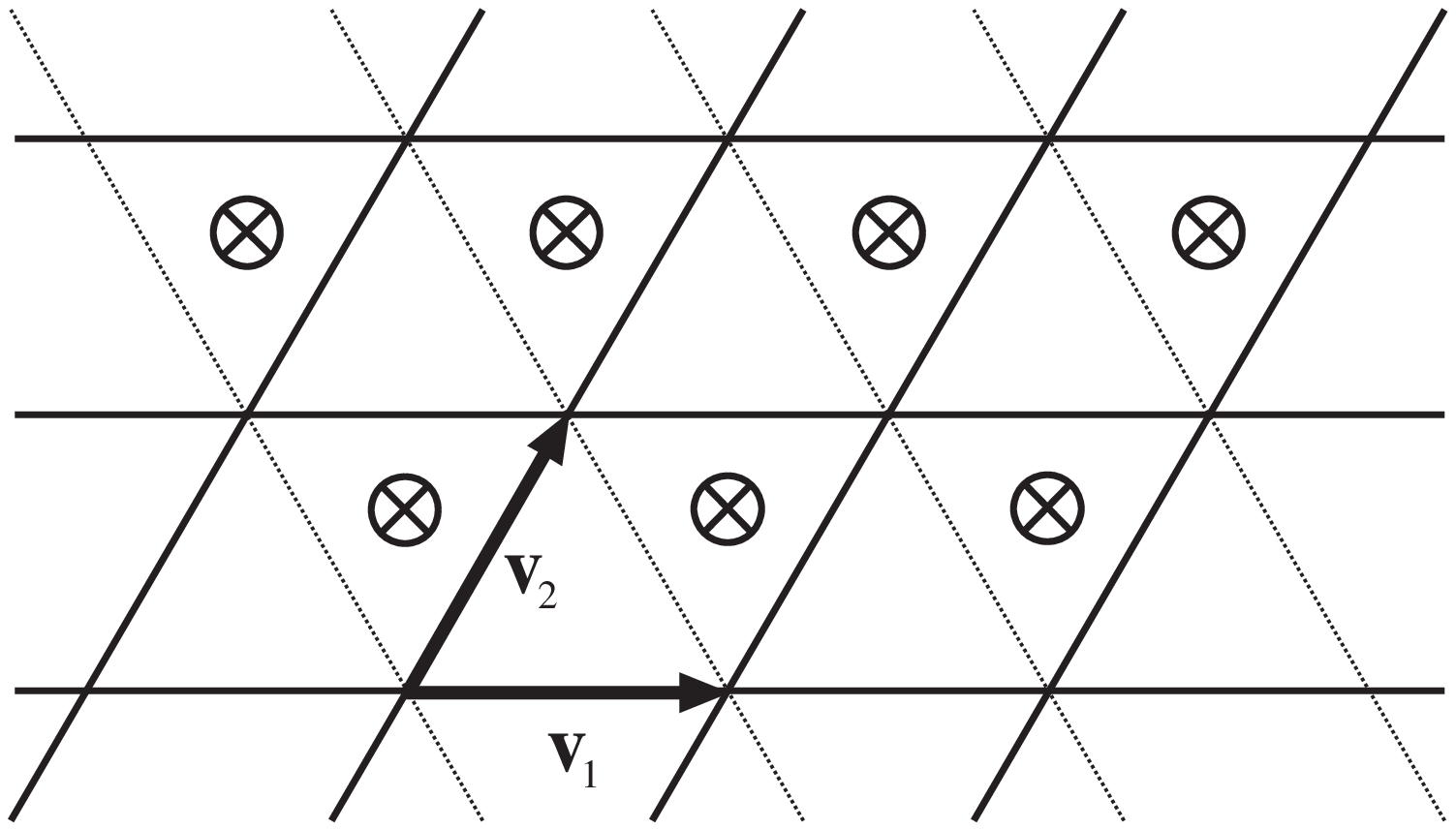}} \\[1cm]
FIG. 3\\[2cm]
\makebox{\epsfbox{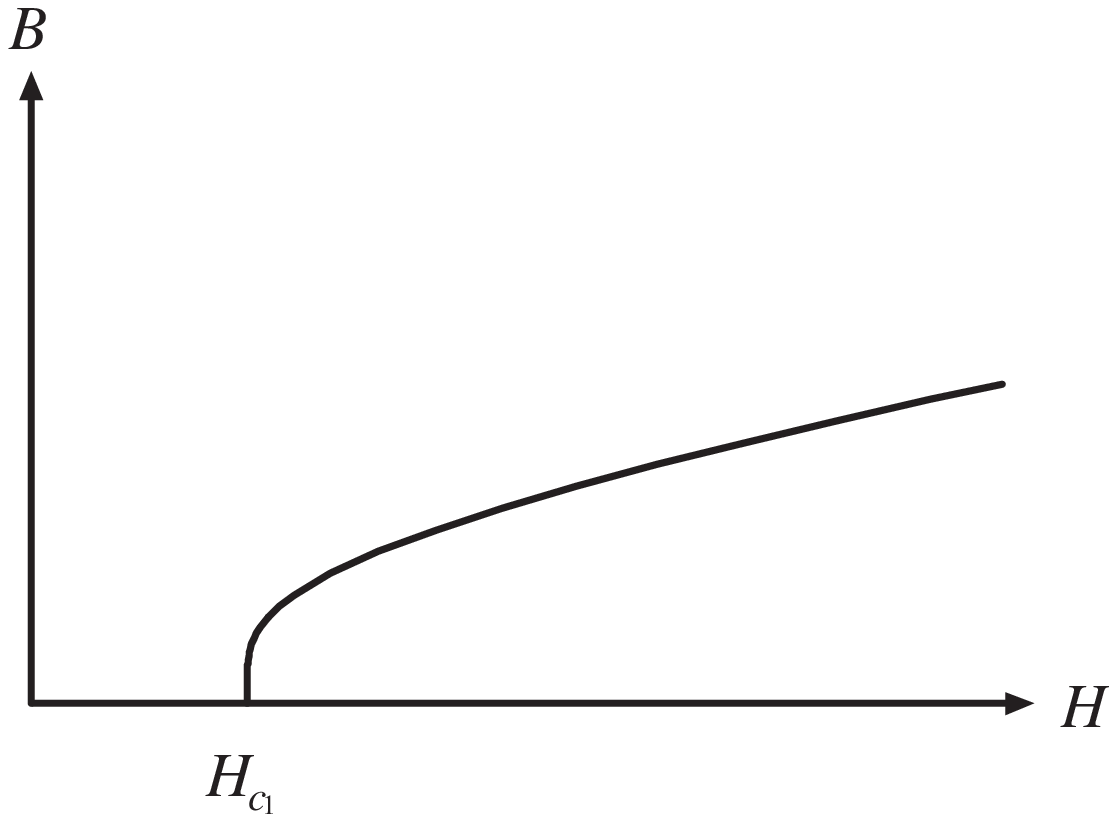}} \\[1cm]
FIG. 4\\[2cm]
\makebox{\epsfbox{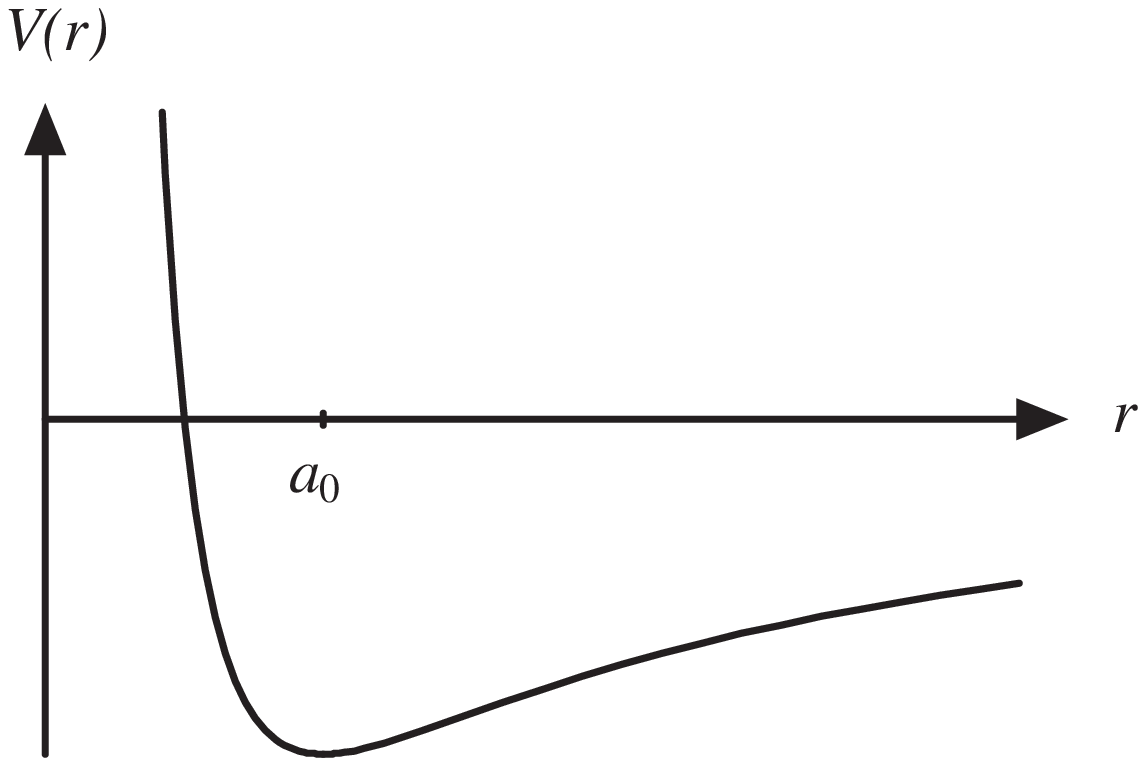}} \\[1cm]
FIG. 5\\[2cm]
\makebox{\epsfbox{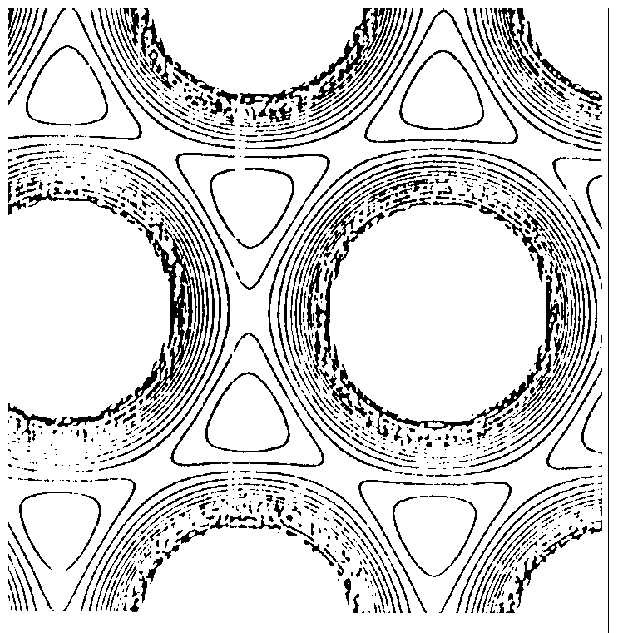}\hspace{1cm}\epsfbox{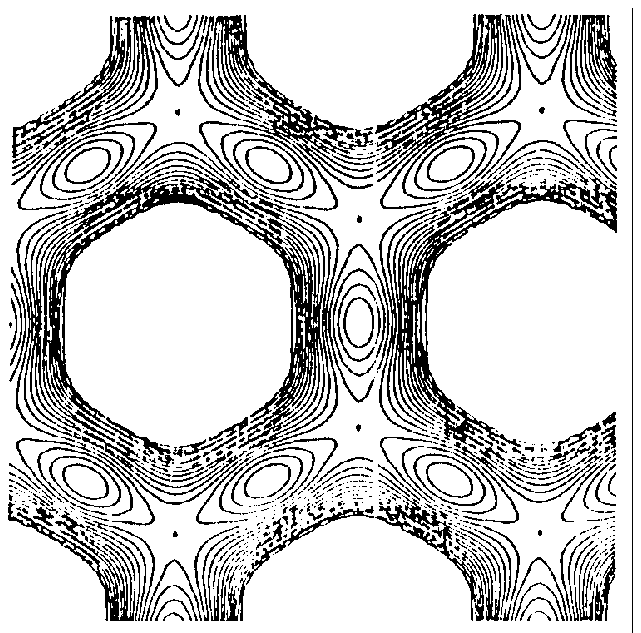}} \\[1cm]
FIG. 6\\[2cm]
\makebox{\epsfbox{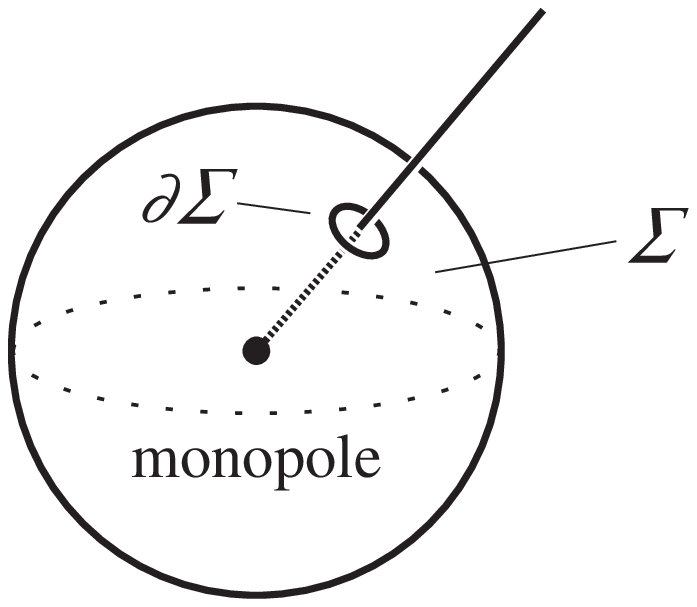}} \\[1cm]
FIG. 7\\[2cm]
\makebox{\epsfbox{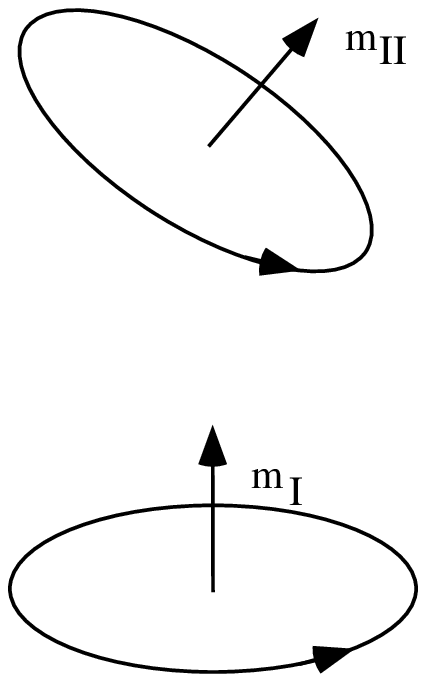}} \\[1cm]
FIG. 8\\[2cm]
\makebox{\epsfbox{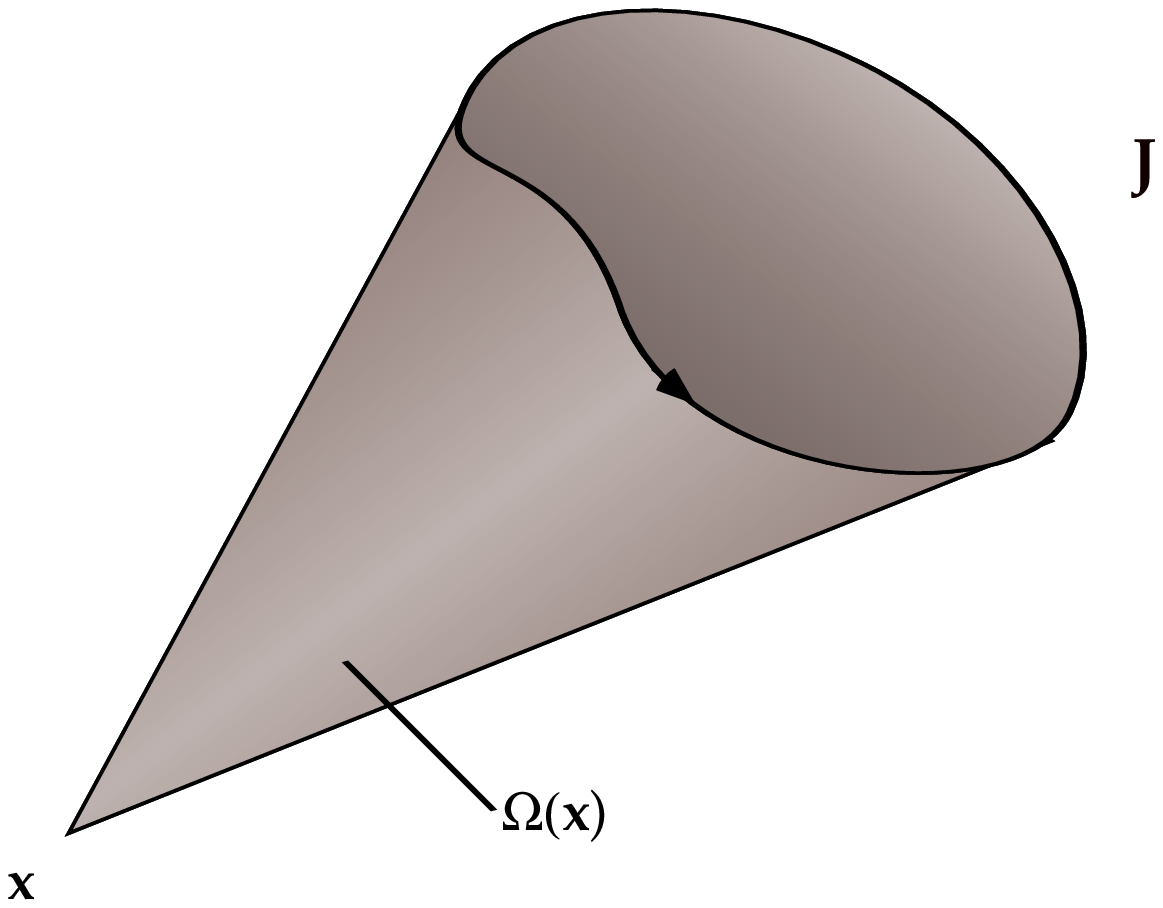}} \\[1cm]
FIG. 9\\[2cm]
\makebox{\epsfbox{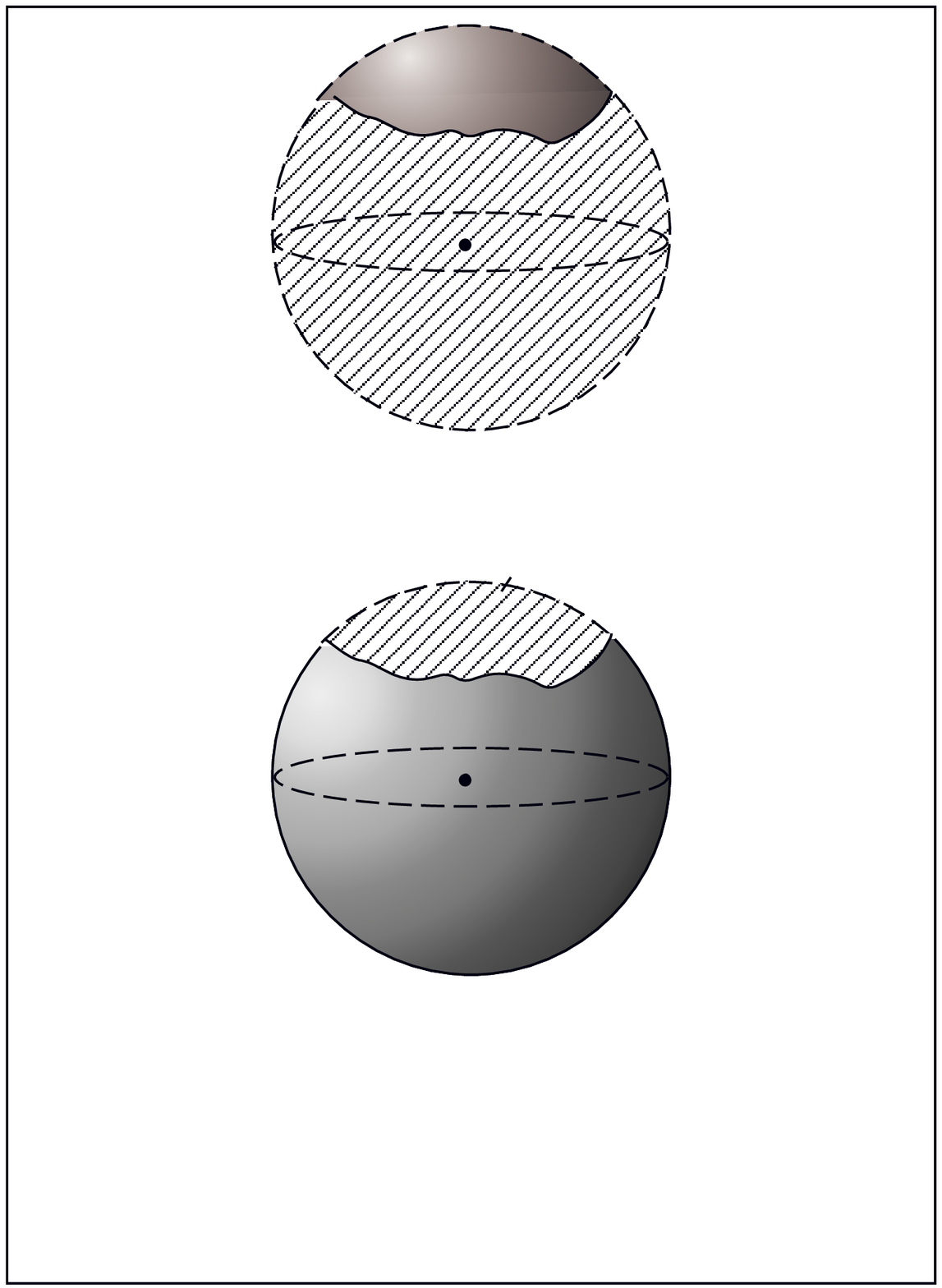}} \\[1cm]
FIG. 10\\[2cm]
\makebox{\epsfbox{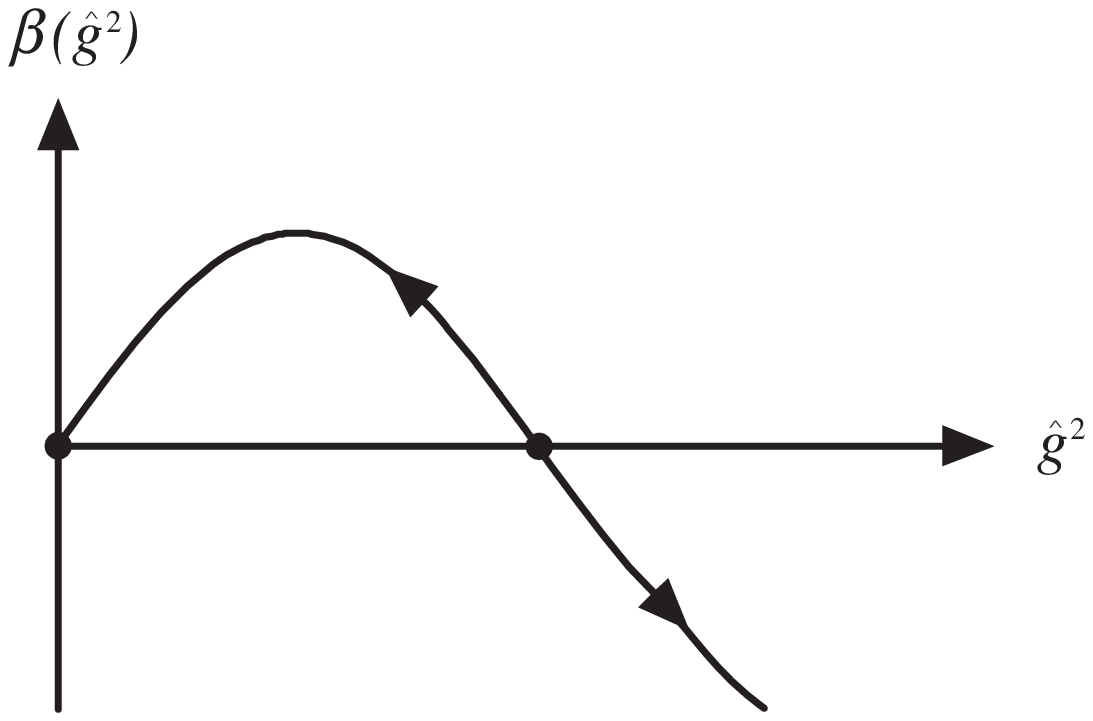}} \\[1cm]
FIG. 11\\[2cm]
\makebox{\epsfbox{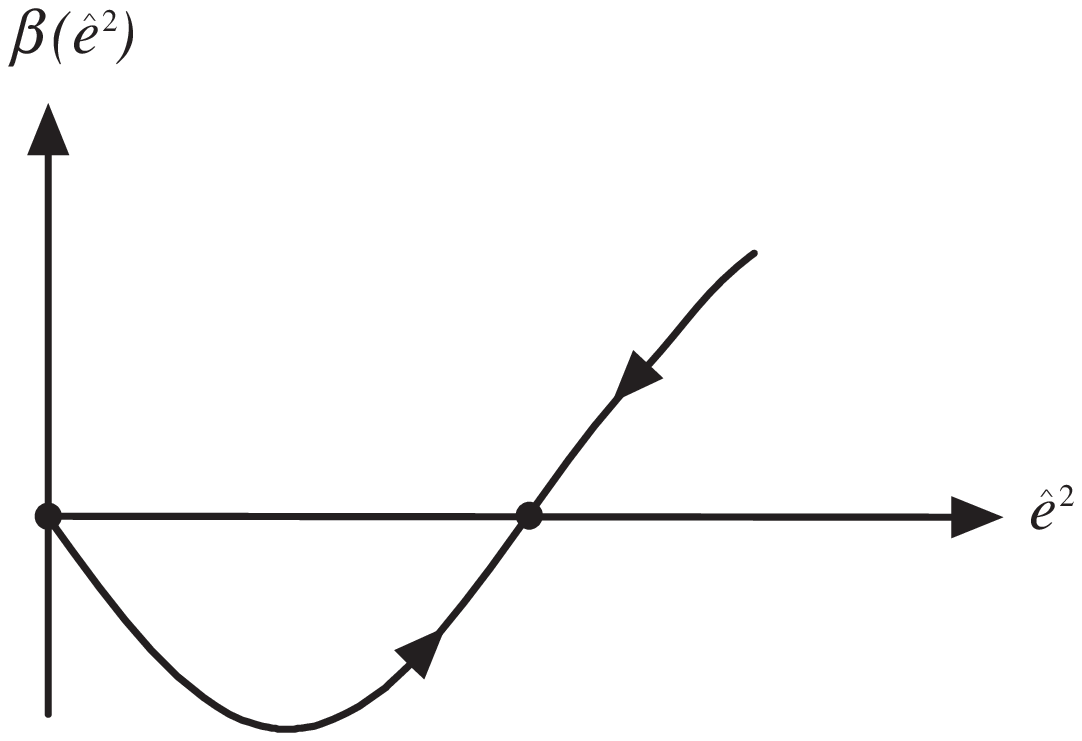}} \\[1cm]
FIG. 12
\end{center}

\end{document}